\documentclass[journal=jacsat,manuscript=article]{achemso}

\usepackage[version=3]{mhchem} 
\usepackage{booktabs}
\usepackage{tabu}
\usepackage{titlesec}
\graphicspath{ {figures/} }
\usepackage{array}
\usepackage{caption}
\usepackage{subcaption}
\usepackage{multirow}
\usepackage{hyperref}
\usepackage{xcolor}
\usepackage{pdfpages}

\author{Audrey Ngambia}
\affiliation[University of Edinburgh, School of Chemistry]
{School of Chemistry, University of Edinburgh, Joseph Black Building, David Brewster Road, EH9 3FJ, United Kingdom}

\author{Ondřej Mašek}
\affiliation [University of Edinburgh, School of GeoSci]
{UK Biochar Research Center, School of GeoScience, University of Edinburgh, Alexander Crum Brown Road, EH9 3FF, United Kingdom}

\author{Valentina Erastova}
\email{valentina.erastova@ed.ac.uk}
\affiliation[University of Edinburgh, School of Chemistry]
{School of Chemistry, University of Edinburgh, Joseph Black Building, David Brewster Road, EH9 3FJ, United Kingdom}

\title[Porous biochar models]{Development of biochar molecular models with controlled porosity}

\abbreviations{MD - molecular dynamics, VA - virtual atom, BSU - basic structural unit}


\begin{document}


\begin{abstract}

Biochars are an exciting class of environmental materials with a wide range of applications, including carbon storage and sequestration, soil enhancement, and pollution remediation.
However, the limited knowledge of their molecular structures and compositions and the lack of comprehensive understanding of the relationship between these structures and biochars' diverse functionality, is hindering advancements in their development.
In this work, we further advance the approach, first introduced by Wood \emph{et al.} (2023), to constructing biochar molecular models; and now include control of microporosity (pores $<$ 2 nm size) within the developed models.
We construct biochar models representative of woody biochars which are experimentally produced at 600 -- 650 °C highest heating temperatures. Our models reproduce experimental H/C and O/C atomic ratios, percentage aromatic carbon, true density, cumulative porosity, and pore size distribution. 
The development of microporous biochar molecular models allows us to identify the importance of chemical structures involved in the assembly of biochar materials, and describe the relationship between these structures and obtained micropores. 
To facilitate other researchers integrating our approach into their work, we detail the steps taken, including the tests and reasons for each decision, in the construction of the biochar models.
Furthermore, we share our developed molecular models in a format that can be easily integrated into other group's work in the form of molecular dynamics simulations.

\end{abstract}

\section{Introduction}

\subsection{Biochar applications and characterisation challenges}

Biochar is a porous carbon-rich material generated from the pyrolysis of biomass.\cite{lehmann2007biochar,tripathi2016effect} 
Biochar has found multiple environmental applications and is an excellent material for carbon storage,\cite{weber2018properties} soil amendment,\cite{o2018biochar, qian2019screening,shaaban2018concise} and adsorption of various pollutants.\cite{xiang2020biochar, wang2020recent} 
What makes biochar materials so versatile for this vast array of applications is their high surface area and the diversity of surface-exposed functional groups. 
For a given biochar, these properties are connected to the starting biomass material (feedstock) and the pyrolysis conditions employed during its production.\cite{janu2021biochar} 
Furthermore, a biochar material can undergo additional pre- or post-treatments to further enhance its properties. 
Whereas one would wish to produce biochars with tailored properties for a bespoke application, the reality is that the process of producing these materials is often left to serendipity, relying on trial and error in the laboratory. Even when the production of biochars is combined with extensive and careful experimental analysis, the connection between biochars' properties and their structure/composition remains, at best, system-specific if not inexistent. 
The current lack of knowledge of what could be coined structure-function relationships in biochars is due to the intrinsic molecular complexity of these materials,\cite{palmer2012atomistic} which is further exacerbated by the variability induced by the precise conditions during biochar production. 
Overall, the lack of knowledge of structure-function relationships has been identified as a major limiting factor in the development and application of these environmentally friendly and sustainable materials like biochars.\cite{lu2020application,ranganathan2017generation,palmer2009detailed}

In this context, molecular modelling offers a transformative strategy for uncovering structure-property relationships and identifying the key interfacial properties in biochars by connecting experimental observables to molecular structures and compositions. 
Molecular dynamics simulations benefit from a strong track record of successful application in areas such as drug discovery,\cite{lin2022applications} biochemistry,\cite{macuglia2022emergence} and material modelling,\cite{evans2017application,tian2018densification,he2016molecular} but surprisingly have not yet been exploited to their true potential for the study and design of biochars. 
The reason for this apparent lack of modelling is not due to the methodology itself but rather to the lack of input structures. A good simulation, indeed, starts from a good input, and biochars being complex amorphous materials makes the generation of an initial representative model structure highly challenging. Such a challenge is evidenced by the very limited number of molecular modelling studies of biochars in the literature. For the few published computational works on biochars, crude approximations are employed, such as representing a biochar as a single benzene-ring disk,\cite{mrozik2021valorisation} a functionalized graphitic flake,\cite{feng2021functionalized} or a graphitic slit pore.\cite{wu2022hierarchically} Such approximations make these models nonrepresentative of the true nature of biochar materials, hampering any connections with experimental observables and, more critically, the determination of structure-function relations to support biochar material development in the laboratory.


\subsection{Biochar properties and structures at the nanoscale}

Biochar materials are generated from different types of biomasses. Among biomass materials, plant residues from agriculture, forestry, or food production are most frequently used thanks to their availability and strict regulations on other biomass sources \cite{meyer2017biochar}. The yield and properties of biochar are highly dependent on the biomass composition and production conditions: the pyrolysis temperature, the heating rate, and the processing time. These conditions affect the physical and chemical properties of the final material, such as surface area, pore volumes, pore sizes and surface-exposed functional groups.\cite{weber2018properties}
As the biomass is heated in the absence of oxygen, processes of dehydration, molecular structure degradation and carbonisation will affect the morphology of the produced biochar.
The porous structures in the biochar material originate from the inherent porous structure of the starting biomass, i.e., macropores $>$ 50 nm in diameter, and from escaping volatile compounds during pyrolysis, i.e., micropores ($<$ 2 nm) and mesopores (2 -- 50 nm).\cite{rouquerol1994characterization} 
Some pores may remain closed, i.e., inaccessible to solvent or gas molecules; therefore, increasing the number and volume of closed pores will not contribute to changes in measurable surface area but will reduce biochar density. However, increasing the number of open pores will increase the surface area of biochar, i.e., the biochars produced at higher pyrolysis temperatures are expected to have larger surface areas, resulting from higher porosity, than those of their lower pyrolysis temperature counterparts \cite{brewer2014new}.
Overall, a biochar material will feature an array of pore sizes that can be presented as a pore size distribution with a corresponding cumulative pore volume.\cite{maziarka2021you,muzyka2023various,das2021compositional,lehmann2007biochar} 
Accordingly, biochar materials with higher pore volumes in the micropore and mesopore regions will contribute the most to the high surface area.
Biochar density can be described by the \emph{bulk} density, a mass of dry material per volume, which includes the volume occupied by pores, and the \emph{true} density, a mass per volume of condensed material only. While, ideally, the latter should not include closed pores, there is no experimental capability to exclude those from measurements of true density, which is defined by the gas molecule access to the pores. However, the effect of the pores is minimal, and the changing porosity of biochar does not directly influence the true density.\cite{weber2018properties,brewer2014new}
 
The chemical structure of biochar is comprised predominantly of carbon, hydrogen, and some heteroatoms (oxygen, nitrogen, sulphur and phosphorus). These heteroatoms form functional groups and create chemical heterogeneity in biochar.
The chemical and structural changes in the biomass during its thermal decomposition towards biochar, result in the detachment of the functional groups and their removal as volatile compounds. This leads to the overall decrease in the H/C and O/C ratios and the overall reduction in the number and variety of functional groups, as the pyrolysis temperature increases.
This is also reflected in the decrease in the less thermally stable acidic oxygen-rich groups (e.g., carboxylic groups), and the increase in the basic oxygen-rich groups (e.g., pyrone, quinone, chromene)\cite{lehmann2007biochar,tomczyk2020biochar}. The simultaneous condensation and graphitisation produce aromatic carbon structures, formed by both planar 6-membered rings and some curved 5- and 7-membered 
rings.\cite{wood2023biochars_I,weber2018properties,lehmann2007biochar}
Such aromatic structures, formed during pyrolysis, indicate the thermal stability of the material and are important when considering long-term biochar stability and functionality.\cite{weber2018properties}
The microscopic aromatic structures of biochar are classified into two main phases: \emph{amorphous} -- randomly stacked cross-linked polycondensed aromatic rings, and \emph{crystalline} -- co-aligned graphitic structures formed by the aromatic sheets.\cite{franklin1951crystallite} 
At the molecular level, the aromatic structures can be characterised by the proportion of aromatic carbon in the structure, termed \emph{aromaticity index}; and the size of the structure formed by the fused aromatic carbon rings, termed \emph{aromatic domain size}.\cite{mcbeath2009variation} 


\subsection{Molecular modelling of biochar and structurally similar materials}

While molecular modelling comprises a range of methodologies allowing the study of systems and processes at various levels of resolution, the system sizes and time scales necessary to describe processes involving biochar materials exclude the use of quantum mechanics calculations, at least during the initial steps of developing the material structures. 
While in the future it may become necessary to include hybrid methodologies, such as QM/MM (quantum mechanics/molecular mechanics) or MD-to-DFT (molecular dynamics to density functional theory) pipeline\cite{goldie2022identification}, we have not yet reached this stage and, therefore, here we focus on the discussion of molecular mechanics approaches only.

Molecular mechanics force fields describe the molecular system as a set of atoms, represented as spheres with a given softness and a fixed charge, connected by bonds, often approximated to a harmonic potential with a given equilibrium length and stiffness, and a set of angles and dihedrals, also defined by equilibrium angle values with a given stiffness or periodicity, respectively.
The interactions between atoms lead to the multidimensional potential energy landscape, which describes the system. In the case of molecular dynamics (MD) simulations, the molecular system evolves on this potential energy landscape following the laws of classical Newtonian mechanics, driven by the balance of thermal motions. In the case of Monte Carlo, the equilibrium statistical ensembles are sampled without the need for dynamics.

The first step in molecular modelling is the generation of input molecular structures. Knowledge of this molecular structure is a limiting step in biochar research, both in experimental and modelling studies. Actually, the problem is so pertinent that it becomes the focus point of much research. Nevertheless, we should not forget that the produced model itself should not be the end goal, but a step to further enable studies of the material for its applications, such as those in environmental settings, that revolve around the interactions between biochar's surfaces with various liquids, gases, or ions.

Although there are very few modelling studies of biochar to date,\cite{franklin1951crystallite,mandal2021sorption,zhou2022adsorption} there is a wealth of work dedicated to structurally similar materials such as amorphous carbon, \cite{farmahini2015hybrid,opletal2002hybrid,opletal2005structure,nguyen2008new} kerogen, \cite{ungerer2015molecular,wang2021molecular,liu2021molecular} and coal \cite{yan2021molecular,marzec1997new, mathews2012molecular}. We will base our further discussion on the advances in modelling these materials to learn and transfer knowledge into our work.

Fundamentally, when attempting to describe a molecular system, there are two approaches: the \emph{top-down} and the \emph{bottom-up}. 
The top-down approach starts with experimental data of the bulk material properties that drive the evolution of the chemical structures within the modelled system to fit this experimental data. 
In contrast, the bottom-up approach starts with the defined chemical structures, which then self-assemble into a condensed material, which exhibits the bulk material properties validated against the experimental measures. 


\subsubsection{Top-down approaches}

The top-down approaches allow for the evolution of chemical structures toward the defined properties. Since classical force field methods rely on first defining the molecule, they are not suitable for this approach. To this end, if using an atomistic description, one would typically resort to the bond-order-based “reactive” force fields.\cite{senftle2016reaxff} 
Here, the system is initiated by a given number and type of atoms and will evolve through MD simulation at high temperatures to generate the lowest energy molecular structures for this pre-defined density.\cite{de2019transferability,obliger2018poroelasticity}
While appealing at first, there are setbacks to using reactive force fields, that became apparent with their use.
Most importantly, the emergence of unphysical structures and significant sensitivity to simulation parameters indicate the lack of reliability and transferability of these force fields.\cite{de2019transferability}

The integration of methodologies to constrain the evolution of molecular structures to an array of experimentally known parameters produces a higher degree of match to the desired (and targeted) experimental descriptors. In such cases, validating the obtained structures against the properties not used as targets is essential. Overall, these methods rely on the use of reverse Monte Carlo \cite{mcgreevy1988reverse} that adjusts the model to match predefined experimental parameters (e.g., pair-distribution functions, structure factors).
Combining reverse Monte Carlo with MD further improves the ability to generate physically plausible structures.\cite{farmahini2015hybrid}

Another top-down approach is to model, or mimic, the chemical process that generates the final structure from a well-known starting material with known by-products.\cite{gelb_2009,shi2008mimetic,mi2009elastic} For example, this approach has been applied to model the pyrolysis of cellulose.\cite{zheng2016initial} While certainly a curious application, in addition to impediments of using reactive force fields described above, this approach has a significant limitation of attainable timescales, where pyrolysis heating rates are in order of 10$^4$ K per ns. This highlights the extremely high computational expense associated with the use of reactive force field methods, limiting their applicability to large systems, such as biochars. For further discussion on this topic, we recommend the review by Obliger \emph{et al.}\cite{obliger2023development} that discusses the most recent advances in reactive force field methodology applied to kerogen research. Generally, the structures produced through reactive force fields are periodic in 3D, which means that the system only represents the bulk material. Therefore, this approach does not accommodate the use of developed models for high-throughput studies of processes at the interface, such as gas or liquid adsorption.


\subsubsection{Bottom-up approaches}

The bottom-up approaches rely on users' chemical knowledge to construct molecular structures that represent the material. While, ideally, a chemical structure is already known from experimental data, which is the case for areas such as drug design, protein modelling, and polymer materials, where MD simulations have become commonplace, this approach is guesswork otherwise. 
When no defined structures are available, one must begin collating the available chemical knowledge, such as the elemental composition, functional group information (e.g., from NMR, XPS, S-XANES, FTIR), or aromatic domain sizes (e.g., from NMR, XRD) and surface area (e.g., BET). The carbon material models are then constructed by finite islands of polycondensed aromatic rings with attached functionalised and nonfunctionalized aliphatic chains. 
The resulting structures are then simulated with MD to allow for self-interaction and self-assembly and then validated through the system's emerged physical properties (e.g., density, viscosity, structural factors). This approach was successfully implemented by Urgerer \emph{et al.} and allowed the construction of molecular models of kerogens with different degrees of maturity.\cite{ungerer2015molecular} Similarly, Boek \emph{et al.} developed a computer algorithm to generate molecular representations of asphaltenes based on experimental data.\cite{boek2009quantitative}
While the creation of molecular models is surely a time-consuming task, it only needs to be done once; and then the models can be used for various applications. This is evident by the popularity of both aforementioned models in further molecular simulation studies. 

While kerogen and asphaltenes are comparable to biochar, these are still distinct materials. 
Biochar materials are solids, formed by heavier molecular structures with large aromatic domains, and models of even the most mature kerogens are still not representative of biochars.
More recently, Wood \emph{et al.} developed models for a range of biochars derived from woody biomass.\cite{wood2023biochars_II} The molecular models were based on the elemental composition and functional group information, and the condensed system reproduced the true density and morphology of woody biochars. Although the approach implemented by Wood \emph{et al.} was able to generate models that reproduce chemical and physical properties, it did not account for porosity, created by the escape of volatile gases during pyrolysis.


\subsubsection{Inclusion of microporosity into carbonaceous material models}

Probably the simplest and most widely used method for the inclusion of pores in carbons is the use of a slit-like pore model.\cite{lastoskie1993pore, palmer2012atomistic}
Such a model is made of two parallel graphitic sheets separated by a width corresponding to the pore size. 
It is clear that the pores created by the slit-like pore method are very simplistic in geometry, which fails to account for the irregular morphologies of pores in nanoporous amorphous materials.
Other strategies have involved deleting atoms to create a void,\cite{williams2016new} or, instead, the insertion of dummy particles during the condensation step of the material simulation.\cite{collell2014molecular,zhou2016Novel}
The atom deletion method could result in a modified and unstable molecular structure due to the breaking of bonds where the pore is created.\cite{zhou2016Novel}
Dummy particle insertion methods have so far produced more realistic pore structures that capture heterogeneous morphologies of carbonaceous materials.\cite{collell2014molecular,zhou2016Novel}

\hfill

In this work, we expand on the iterative approach for the construction of biochar models, introduced in the work of Wood \emph{et al.}, \cite{wood2023biochars_II} and add the capability of controlled porosity. 
We focus on biochars produced at 600 -- 650 °C from softwood pellets, as these materials provide a variety of pore sizes, ideal for the purpose of this study.
Our approach incorporates the use of building blocks to construct the biochar models: the carbon matrix of biochar is described by molecular building blocks, while the controlled porosity is incorporated into the carbon matrix through virtual atoms (VAs). The philosophy behind this approach is detailed in Section \ref{Ch:Philos}.
In Section \ref{Ch:VAs}, we evaluate the parameters of VAs necessary to introduce the desired pore sizes into the condensed carbon materials. 
We then present the biochar models (Section \ref{Ch:PureBC}), constructed only using molecular building blocks, as in the original work by Wood \emph{et al.} protocol, and discuss the structures and properties of these models.
In the following Section \ref{Ch:BCVA}, we integrate VAs into the construction of the models to obtain microporous biochar models.
This allows us to draw attention to the importance of certain structural moieties within biochar material for its stability as a whole.
With this work, we present an approach for the construction of biochar models that are truthfully representative of the experimental counterparts, and we share the structures of microporous woody biochars ready to be used in future studies in the biochar community. 
Our models are freely available from our GitHub page: \url{https://github.com/Erastova-group/Porous_Biochars_Models}.

\section{Results and discussion}

\subsection {Principles of the biochar molecular assembly approach}
\label{Ch:Philos}

\begin{figure}[htpb!]
    \centering
    \includegraphics[width=1\linewidth]{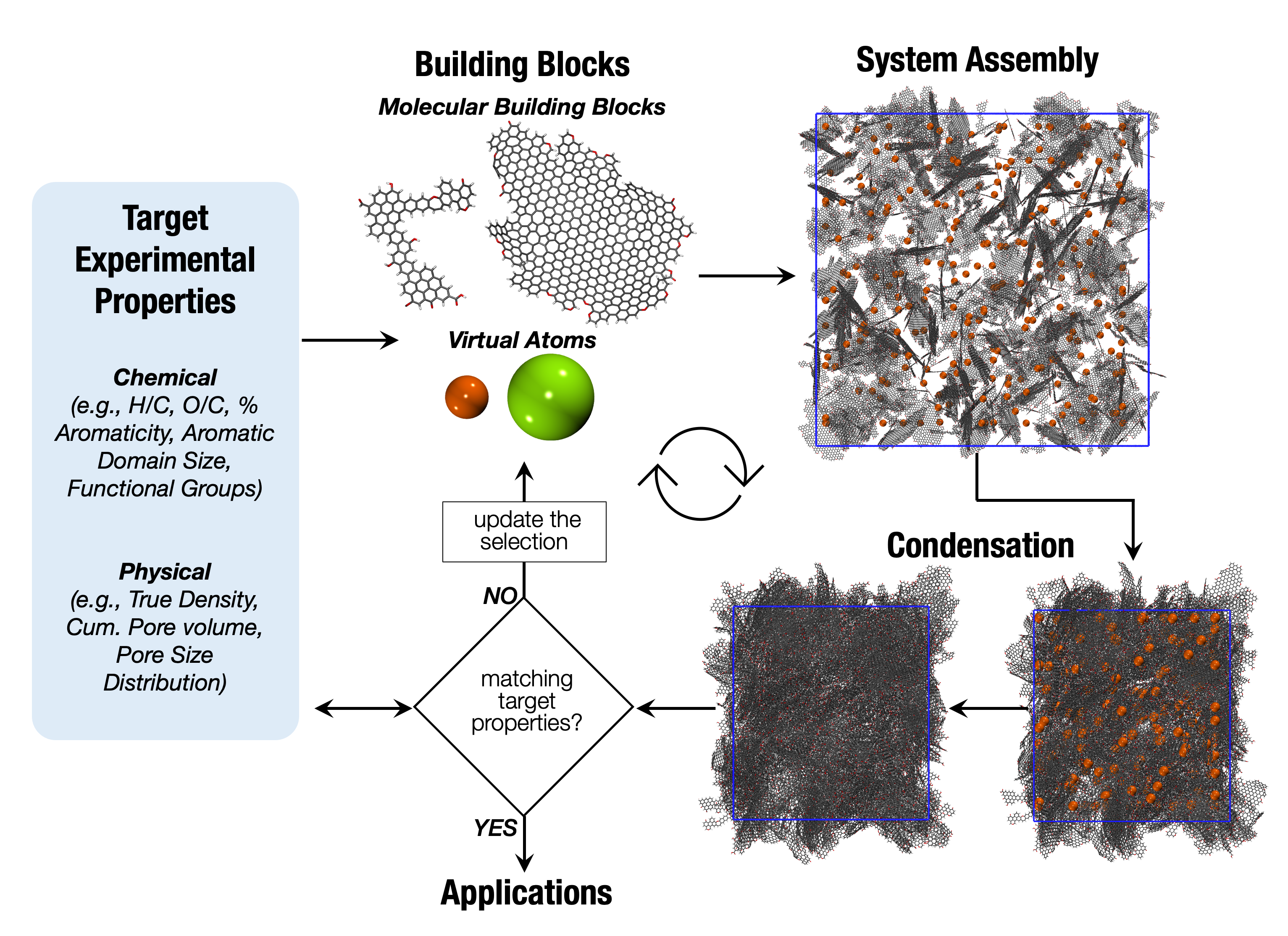}
    \caption{Schematic representation of the iterative protocol for the set-up of realistic biochar molecular models for molecular dynamics simulations, guided by the experimental data.}
    \label{fig:pipeline}
\end{figure}

The foundations of this work are rooted in the iterative approach, first introduced by Wood \emph{et al.} \cite{wood2023biochars_I, wood2023biochars_II}
Our expanded approach is presented in Figure \ref{fig:pipeline} (further details can be found in the methods section \ref{Ch:Methods}), and follows the steps:
\begin{enumerate}
    \item Identification of \textbf{target experimental properties} for the biochar of interest;
    \item Selection of an array of \textbf{building blocks} to create a composite structure that exhibits the specified \textbf{chemical properties};
    \item Step-wise annealing of the molecular system to form an \textbf{equilibrated condensed bulk material};
    \item Determination of the \textbf{physical properties} of the system and validation against the target experimental properties;
    \item If the system matches the targets, the production of a \textbf{surface-exposed model of the biochar}; 
    else, return to \emph{Step 2} with an alteration of the selected building blocks.
\end{enumerate}

This approach is based on the use of a collection of experimental data that would provide a comprehensive description of biochar in terms of both chemical and structural (or physical) properties.  It is essential to separate the two sets of properties, as we will now explain.

Chemical descriptors (e.g., elemental ratios, functional groups, aromatic domain sizes) will remain fixed during a molecular dynamics (MD) simulation and, therefore, are ideal for choosing the make-up of the initial "best guess" system in \emph{Step 2}.
Meanwhile, structural descriptors (e.g., skeletal density, pore size distribution) will emerge from the interactions between the building blocks comprising the "best guess" system during the condensation simulations. Therefore, these descriptors are well-suited for the validation of the selection made in \emph{Step 2}.

Biochars are amorphous materials produced from feedstocks of biomass origin, and, as a result, biochar will display an array of properties, each with a degree of variability. And even within the same material produced in one batch, each sub-sample will never be identical to its neighbour. 
Therefore, where possible, biochars are best described by an assembly of properties, each given as a mean and a standard deviation rather than by a single quantity or measurement. 
Knowledge of deviations in the experimental data further assists by informing us on the confidence windows.
If one must use a single measurement, it has to be taken in the context of the average properties of this type of material and account for the trends observed for biochars produced at the given temperature. 
To assist in such scenarios, we have collated information from the UKBRC Charchive (\url{https://www.charchive.org/}), the UC Davis Biochar Database (\url{http://biochar.ucdavis.edu/}) and published works from the family of woody biochars. The dataset can be found on our GitHub page (\url{https://github.com/Erastova-group/Biochar_MolecularModels}) and the trends in the data are discussed in Wood \emph{et al.}\cite{wood2023biochars_I}

Importantly, the molecular system assembled from a selection of building blocks in \emph{Step 2} does not yet represent an intimately interacting molecular matrix -- a condensed material. 
Therefore, in the following \emph{Step 3} we perform an MD simulation: the simulation starts at a high temperature, which allows the building blocks to sample the phase space; the system is then cooled down with a stepwise annealing protocol, allowing the molecular assembly to relax toward an equilibrated system (for methodology see Section \ref{Ch:BC}). 
The concept of an \emph{equilibrated amorphous solid} may be puzzling. Amorphous solids are formed through a glass transition upon cooling from the melt and, as a result, they will be far from thermal equilibrium, i.e., exist in a metastable state. However, as long as mechanical stability is achieved in the material via self-organisation and redistribution of the forces, in the absence of thermal input, the amorphous solids will not undergo further transitions to lower energy configurations (e.g., densification or regional re-crystallisation).\cite{tong2020emergent, lemaitre2018stress} 
That is the equilibrated structure we are seeking to obtain through the MD simulations in \emph{Step 3}. Practically, we must ensure that the annealing protocol is robust and the physical properties of condensed systems are independent of the annealing protocol we have used. This is discussed in Wood \emph{et al.}\cite{wood2023biochars_II}. Furthermore, as already mentioned, each biochar sub-sample will never be identical, it is recommended to perform the MD condensation in a few replicas and use the average of the measured properties.

It should be noted that while some degree of porosity will emerge during MD condensation, experimentally, microporosity is created by escaping volatile compounds, and so the control on this property should be incorporated at this stage, which we accomplish through the inclusion of Virtual Atoms (VAs) at the point of selection of the building blocks \emph{Step 2}. These VAs are large soft massless repulsive Lennard-Jones spheres that represent the escaping volatiles that interact with the condensing matrix of the biochar to form nonspherical pores.

In the following Section \ref{Ch:VAs} we examine the pores produced by a range of VAs in simple pure hydrocarbons (pentadecane, phenol, toluene, nonane, nonanoic acid, and coronene), as those individual hydrocarbons have characteristics functional groups that are collectively inherent in biochar materials.

\subsection{Assessment of virtual atoms for the creation of porosity in simple hydrocarbon materials}
\label{Ch:VAs}

Virtual atoms are massless 'dummy' atoms. Their purpose is to introduce a repulsive potential during the condensation of a biochar model, mimicking the process of escaping volatile compounds during pyrolysis and consequent cooling of the biochar into a solid phase.
For the simplicity of utilisation of the VAs within the GROMACS algorithm, we use a 12-6 Lennard-Jones potential to describe VAs:

\begin{equation}
V_{LJ} = 4 \epsilon \Bigl[\left( \frac{\sigma}{r} \right)^{12} - \left( \frac{\sigma}{r} \right)^{6} \Bigr] ~,
\label{eqn:LJ}
\end{equation}
where $r$ is a distance between atoms, $\epsilon$ is the depth of potential well or \emph{dispersion energy}, and $\sigma$ is the distance where the potential energy $V_{LJ}$ is zero.
As seen from equation \ref{eqn:LJ}, the Lennard-Jones potential consists of two parts: a steep repulsive 12-term, which describes Pauli repulsion at short distances, and a smoother attractive 6-term, which describes London dispersion forces at long distances.

\begin{figure}[htpb!]
    \centering
    \includegraphics[width=1\linewidth]{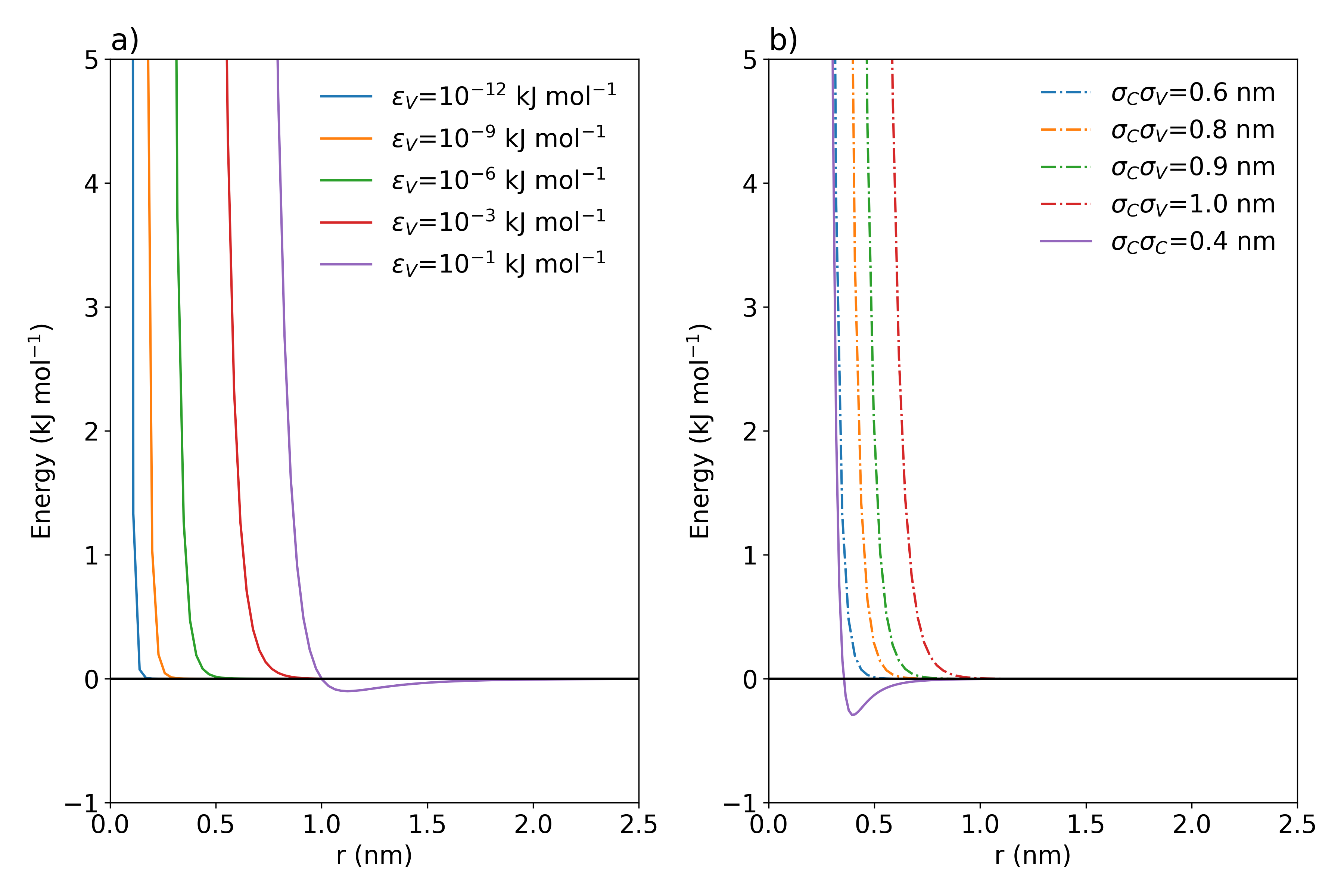}
    \caption{Potential of a) virtual atoms at $\sigma_{V}$ of 1 nm and $\epsilon_{V}$ of 10$^{-12}$, 10$^{-9}$, 10$^{-6}$, 10$^{-3}$ and 10$^{-1}$ kJ mol$^{-1}$ and b) carbon-only potential $\sigma_{C}$$\sigma_{C}$ and carbon mixed with virtual atoms of $\sigma_{C}$$\sigma_{V}$ of 0.6, 0.8, 0.9 nm and 3.0 nm and $\epsilon_{V}$ of 10$^{-6}$ kJ mol$^{-1}$.}
    \label{fig:LJ}
\end{figure}

Figure \ref{fig:LJ}(a) shows the effect $\epsilon_{V}$ has on the potential energy profile while keeping $\sigma_{V}$ fixed at 1 nm.
As $\epsilon_{V}$ increases from $10^{-12}$ kJ mol$^{-1}$ to 0.1 kJ mol$^{-1}$ (in steps of $10^{-12}$, $10^{-9}$, $10^{-6}$, $10^{-3}$ and 0.1 kJ mol$^{-1}$) the slope of the repulsive potential loses steepness, that is, the repulsion becomes softer.
At the same time, the depth of the well also increases, introducing some level of attraction.
To relate the magnitude of the observed energy values, we note that the MD simulation samples thermally accessible states, which are on the order of $k_BT$, i.e., $\sim$ 2.5 kJ mol$^{-1}$ at room temperature.
For further testing, we select VAs with $\epsilon_{V}$ values of $10^{-9}$, $10^{-6}$ and $10^{-3}$ kJ mol$^{-1}$.

In the simulation, the VAs will interact with the atoms that comprise the biochar material. Therefore, we must also consider the effect of combining the two potentials. 
In our simulations, for the organic components we are using the OPLS-AA force field\cite{jorgensen1996development}, which utilises the geometric combination rules:
\begin{equation}
\sigma_{ij} = \sqrt{\sigma_{ii} + \sigma_{jj}}   ~ ,
\label{eqn:comb1}
\end{equation}
\begin{equation}
\epsilon_{ij} = \sqrt{\epsilon_{ii} \epsilon_{jj} }   ~ ,
\label{eqn:comb2}
\end{equation}
where, for two particles $i$ and $j$, each described by their individual $\sigma$ and $\epsilon$, the combined $\sigma_{ij}$ and $\epsilon_{ij}$ are, respectively, the potential well and an inter-particle distance where the potential energy is zero.

The combined potentials that describe the interaction between a VA and an aromatic carbon atom (OPLS-AA parameters $\sigma_{C}$ = 0.355 nm and $\epsilon_{C}$ = 0.293 kJ mol$^{-1}$) are given in Figure \ref{fig:LJ}(b).
While the carbon-only potential (solid line) shows both attractive and repulsive interactions, the combined potentials maintain the soft-repulsive behaviour observed for the given epsilon (Figure \ref{fig:LJ}(a)).  
We have selected VAs with $\sigma_{V}$ of 1.0, 2.2, 3.0 and 4.5 nm for testing with the pure organic systems.

For pure organic systems, we have chosen phenol, toluene, nonane, nonanoic acid, coronene and pentadecane. 
These systems offer a range of carbon structures (i.e., \emph{sp2}, \emph{sp3}, aromatic carbons) and functional groups that cover components which are part of biochar building units, allowing us to access the interactions before moving onto more complicated biochar systems.
The systems consist of compounds which are liquid (nonane, nonanoic acid, toluene and pentadecane) and crystalline (phenol and coronene) at room temperature. 
For liquid systems, the simulation was performed at 300 K in an NPT ensemble, except for pentadecane, which was found to be gel-like at 300 K and, therefore, was annealed to 350 K and then cooled to 300 K.

For the crystalline systems, to ensure that the experimental crystal structures are developed in the simulation, we implemented a stepwise annealing protocol, starting at temperatures above their melting point and then dropping in steps of 25 K per ns and allowing the system to relax for a further 20 ns, by this aiming to prevent hysteresis.
Furthermore, since crystallisation itself produces non-isotropic systems, it is essential to decouple the system scaling on each axis. (See Section \ref{Ch:simple_HC} for full details of simulation protocols.)

For each of the final condensed-phase systems, we calculated densities, reported in the SI Table S1. The densities are in good agreement ($\sim$ 1\% deviation) with the experimental measurements for most of the systems. In the case of coronene and pentadecane, the simulated densities deviate by $\sim$ 10\% from the experimental values. 
The elevated density of the simulated pentadecane system was previously reported by Siu \emph{et al.},\cite{siu2012optimization} and is associated with the formation of gel-like structures at 300 K and a shifted melting point at a higher temperature of 350 K. While this shift in the melting temperature suggests that OPLS-AA parameters should be optimised for better agreement with the experiments, this is not within the scope of this work. 
In the case of coronene, the lower density of the model is associated with the imperfect packing of the coronene molecules during the formation of a needle-like crystal. The crystal growth in the molecular simulation is not a straightforward task, impaired by the time and size limitations of the simulated system.\cite{erastova2012molecular} (Further information on the coronene system can be found in the SI Section S2.)

For each of the pure organic systems, VAs were added in a ratio of 1 VA per 1000 carbon atoms. The VAs tested were combinations of $\sigma_{V}$ $\in$ [1.0, 2.2, 3.0, 4.5] nm and $\epsilon_{V}$ $\in$ [$10^{-9}$, $10^{-6}$, $10^{-3}$] kJ mol$^{-1}$, a total of 12 VAs have been tested. The same simulation protocol was then followed for VA-free systems.

With the increase of $\sigma_{V}$ and $\epsilon_{V}$ of the VAs, the densities of simple hydrocarbon systems decreased (see SI Figure S1). 
The drop in density for all systems is a direct indication of the presence of extra free volume created by the VAs.
As desired, the soft potentials of the VAs allowed the rigid molecular structures to enter within the radius of the VA, creating irregular pores. This also implies that the pore sizes generated by the VAs are considerably smaller than the space defined by their Lennard-Jones parameters.
This was also observed in other studies employing a similar dummy particle approach for shale kerogen\cite{zhou2016Novel,collell2014molecular} and nanoporous carbon.\cite{luo2021virtual}
These observations enable us to establish a dependence between the Lennard-Jones parameters that define the VAs and the resulting pore size (volume and diameter) and roughness (standard deviation of the pore size). To aid the selection of VAs for the next steps of this work, we summarise the VA parameters and pore values in Table \ref{tab:PoreVol}.

\begin{table}[htpb!]
    \centering
    \begin{tabular}{c c c c c } 
          \hline
  VA name&$\sigma_{V}$ (nm) &$\epsilon_{V}$ (kJ mol$^{-1}$)& $V_{pore}$ (nm$^3$)& $r_{pore}$ (nm)\\ 
      \hline
                V10-9 & 1.0 & $10^{-9}$ & 0.17$\pm$0.36 & 0.24$\pm$0.21 \\ 
                V10-6 & 1.0 & $10^{-6}$ & 0.21$\pm$0.18 & 0.35$\pm$0.09 \\ 
                V10-3 & 1.0 & $10^{-3}$ & 0.41$\pm$0.26 & 0.45$\pm$0.08 \\ 
        \hline
                V22-9 & 2.2 & $10^{-9}$ & 0.33$\pm$0.32 & 0.41$\pm$0.11 \\ 
                V22-6 & 2.2 & $10^{-6}$ & 0.65$\pm$0.27 & 0.53$\pm$0.06 \\ 
                V22-3 & 2.2 & $10^{-3}$ & 1.36$\pm$0.26 & 0.69$\pm$0.04 \\ 
        \hline
                V30-9 & 3.0 & $10^{-9}$ & 0.48$\pm$0.39 & 0.47$\pm$0.11 \\ 
                V30-6 & 3.0 & $10^{-6}$ & 1.00$\pm$0.23 & 0.62$\pm$0.05 \\ 
                V30-3 & 3.0 & $10^{-3}$ & 2.30$\pm$0.37 & 0.82$\pm$0.05 \\ 
        \hline
                V45-9 & 4.5 & $10^{-9}$ & 0.76$\pm$0.26 & 0.56$\pm$0.06 \\ 
                V45-6 & 4.5 & $10^{-6}$ & 1.97$\pm$0.39 & 0.78$\pm$0.05 \\ 
                V45-3 & 4.5 & $10^{-3}$ & 5.15$\pm$0.79 & 1.16$\pm$0.06 \\ 
        \hline
    \end{tabular}
\caption{Average pore volume, $V_{pore}$, and pore radius, $r_{pore}$, created by a virtual atom in hydrocarbon systems studied. The standard deviation is indicative of the softness of the pore created.}
\label{tab:PoreVol}
    
\end{table}

\subsection {Development of biochar without added microporosity}
\label{Ch:PureBC}

In this section, we assess the effect of the \emph{basic structural units} (BSUs), the molecular component of the building blocks in \emph{Step 2} of the protocol, on the properties of biochar produced.

Following the protocol, we begin by defining the target properties that describe wood-derived biochar produced at temperature ranges between 600 and 650 °C. Although there is a wealth of published data, discrepancies arise due to a variety of starting materials, even with the same material type, processing and preparation conditions, and limitations from analytical and instrumental techniques. Besides, woody biomass is a structurally complex material, and it is no surprise that the biochar derived from it is highly heterogeneous with several functional groups, collectively contributing to the properties of biochar.
Table \ref{tab:BCtarget} summarises the target chemical and physical properties that describe woody biochars, taken from the data collated by Wood \emph{et al.}\cite{wood2023biochars_I}

\newcolumntype{P}[1]{>{\centering\arraybackslash}p{#1}}

\begin{table}[htpb!]
    \centering    
    \begin{tabular}{P{5em}  P{5em} P{5em} P{5em} P{5em} P{5em} } 
    \hline 
         Biochar &  H/C &  O/C &  \% Aromatic carbon &  True density  (kg m$^{-3}$)  & Cum. pore volume (cm$^3$ g$^{-1}$)\\
         \hline 
         Woody biochar&  0.3$\pm$0.19 &  0.08$\pm$0.06 &  60-90 &  1546$\pm$93  & 0.118-0.175\\ 
         \hline
    \end{tabular}
      \caption{Target properties of woody biochars produced at 600 -- 650 °C, data collected from literature \cite{wood2023biochars_I};  cumulative pore volume of micropores only (pore size $\leq$ 2 nm) for the woody biochars produced between 500 and 700 °C, data taken from \cite{sigmund2017biochar, maziarka2021tailoring}. }
    \label{tab:BCtarget}
\end{table}

\begin{figure}[htpb!]
    \centering
    \subfloat[\textbf{BSU I} \newline H/C: 0.29; O/C: 0.07; \%arom.C: 71, FGs: O-H, C-O-C; ADS: 22.]
    {
        \label{fig:BSUs_I}
        \includegraphics[width=0.4\textwidth]{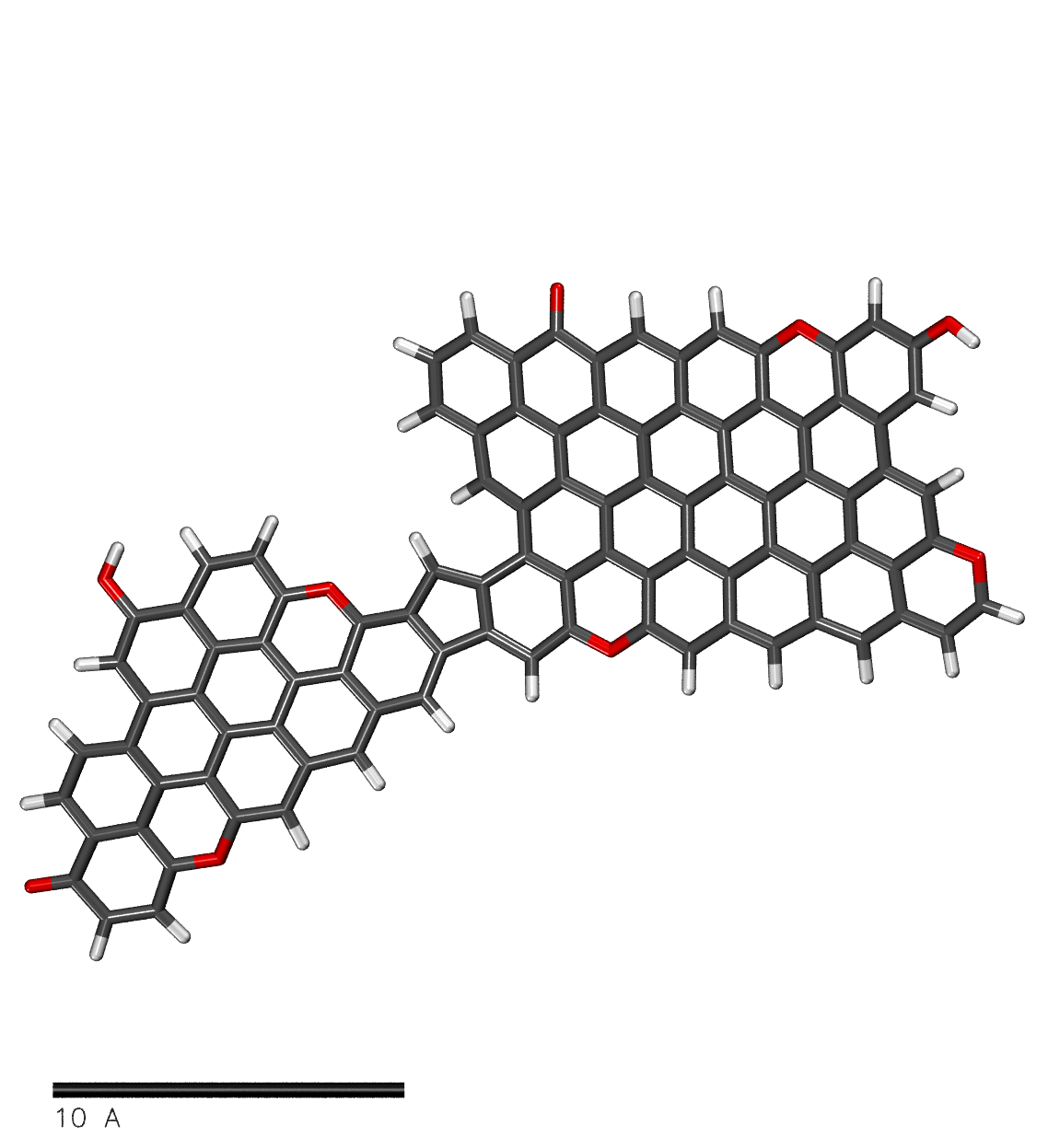}
    }\hfill
    \subfloat[\textbf{BSU II}  \newline H/C: 0.27; O/C: 0.07; \%arom.C: 72; FGs: O-H, COOH, C-O-CH$_3$; ADS: 33.]
    {
        \label{fig:BSUs_II}
        \includegraphics[width=0.4\textwidth]{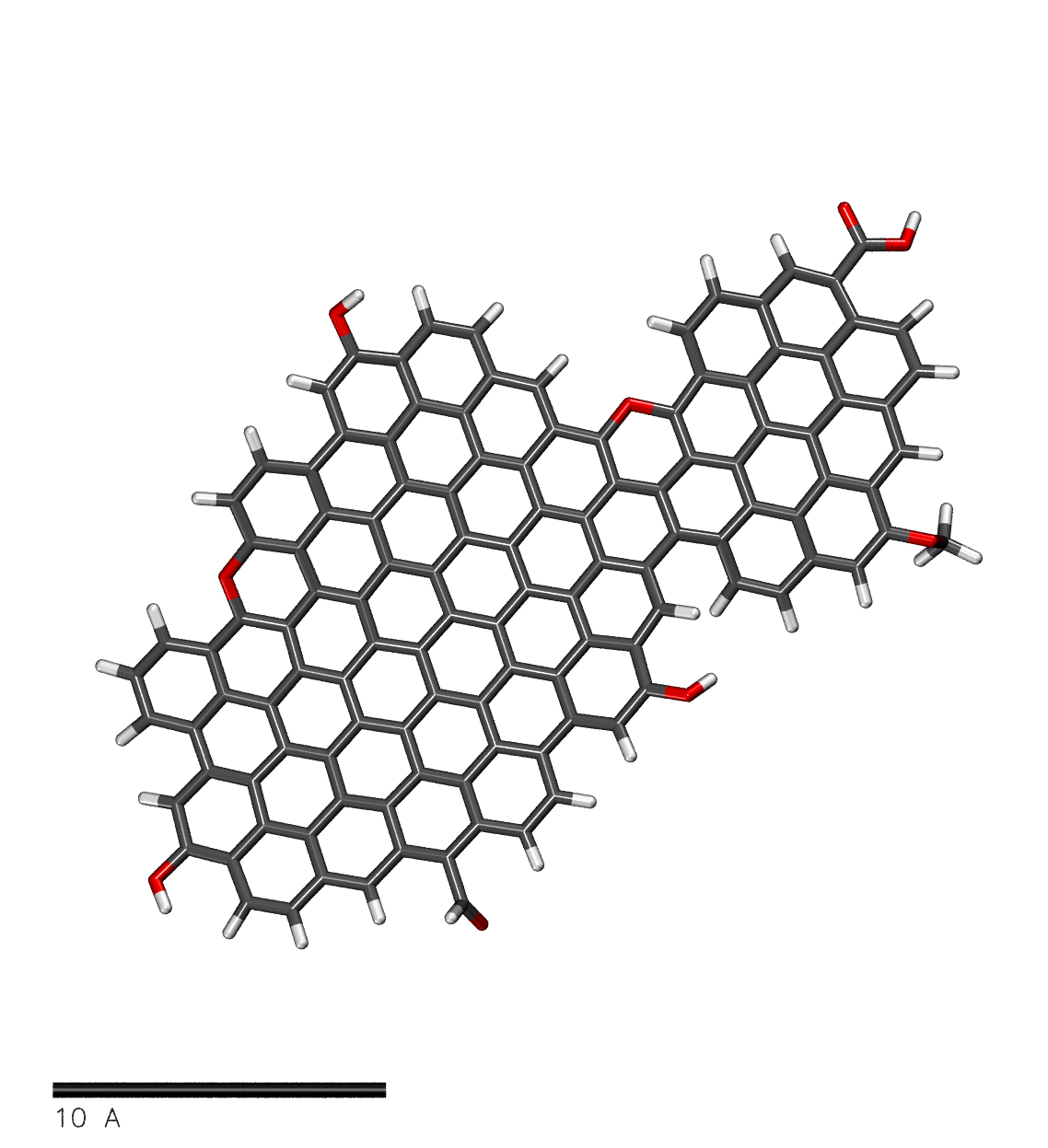}
    }\hfill
    \newline
        \subfloat[\textbf{BSU III}  \newline H/C: 0.41; O/C: 0.11; \%arom.C: 59; FGs: O-H, COOH, C=O; ADS: 9.] 
        {
        \label{fig:BSUs_III}
        \includegraphics[width=0.4\textwidth]{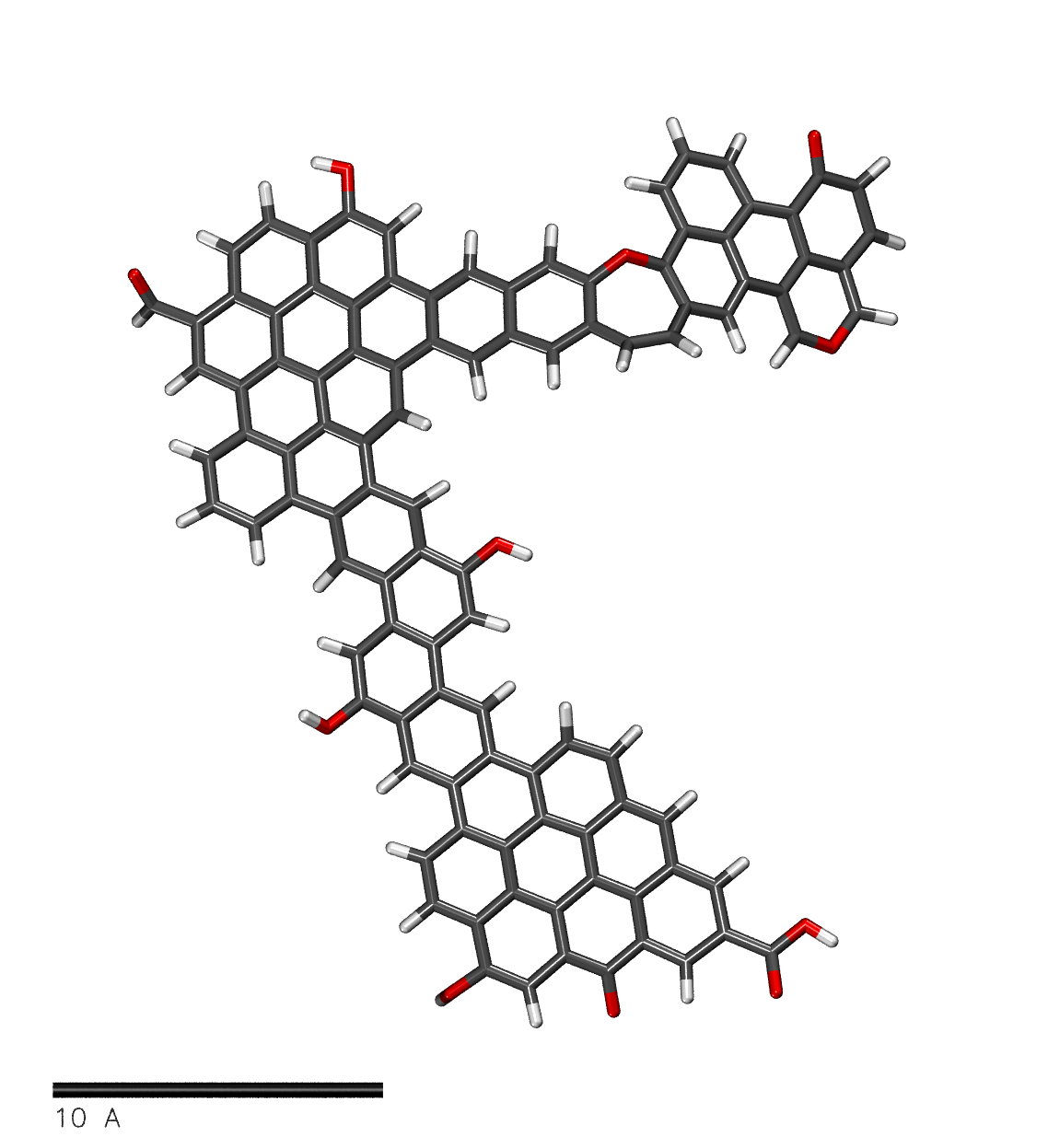}
    }\hfill
    \subfloat[\textbf{BSU IV}  \newline H/C: 0.11; 0/C: 0.03; \%arom.C: 89; FGs: O-H, C-O-C; ADS: 425.] 
    {
        \label{fig:BSUs_IV}
        \includegraphics[width=0.4\textwidth]{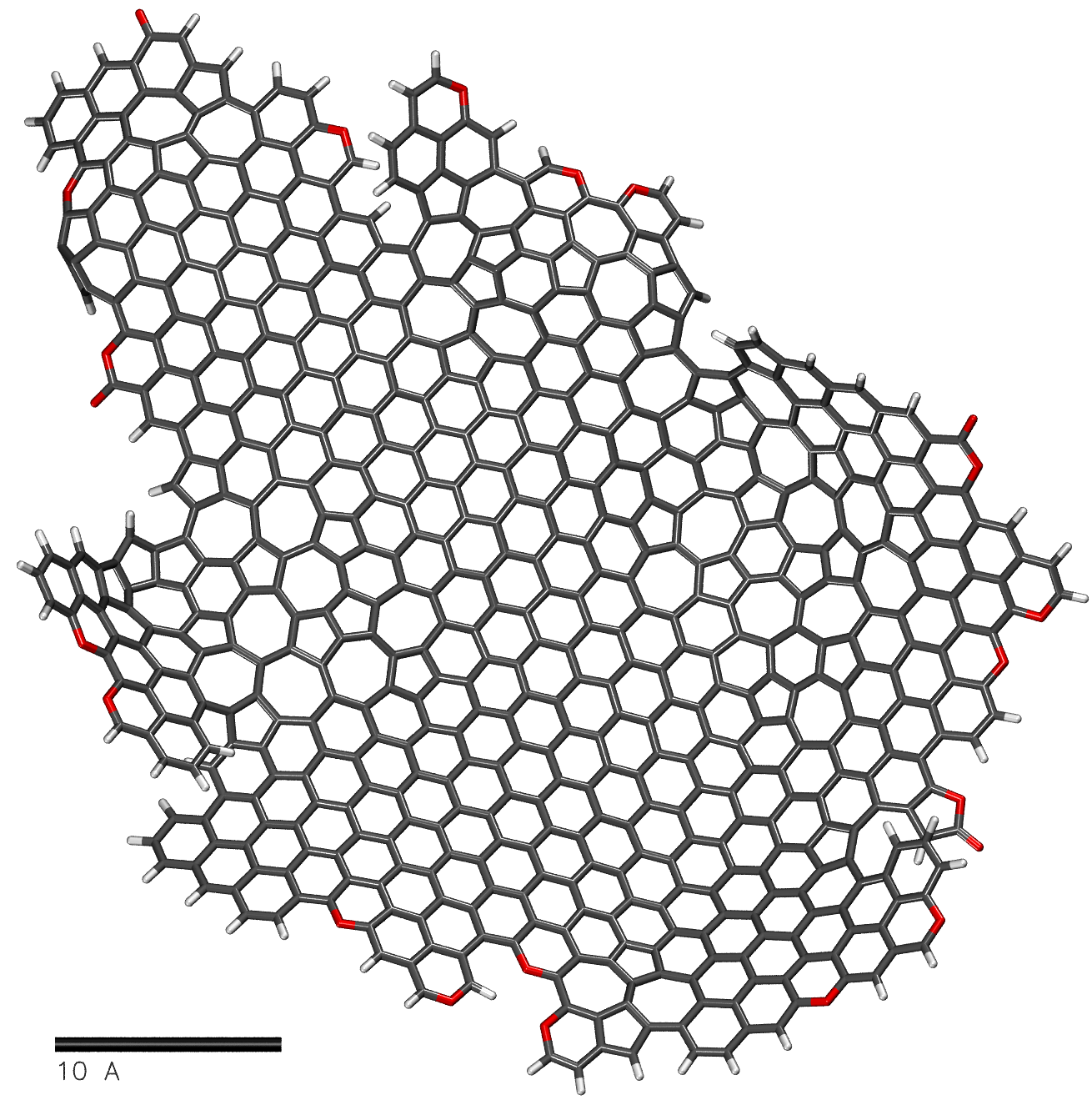}
    }\hfill
    \caption{Basic structural units (BSU) used in this work and their descriptors: H/C and O/C atomic ratios, \% aromatic carbon (\%arom.C), functional groups (FGs) and aromatic domain size (ADS). BSU IV is taken from Wood \emph{et al.}\cite{wood2023biochars_I}. Colours: C - grey, O - red and H - white; scale bar of 1 nm is given.}
    \label{fig:BSUs}
\end{figure}

In the next step of the protocol, we select BSUs to match the chemical description of the target biochar. 
To this end, one may use a variety of structures that collectively result in the desired selected quantities. 
In this work, the four BSUs (I-IV), presented in Fig. \ref{fig:BSUs} are used. We note that these BSUs are not exhaustive, and several BSUs could be constructed to match the same properties. 
The BSUs in this study are chemically composed of carbon, hydrogen and oxygen only. Carbon is in its majority aromatic with 6-membered rings but also features some odd-membered (5 and 7) rings. These odd-membered rings create curvature in planar graphitic sheets and, while present in all biochars, notably arise from the carbonisation of biomass composed of cellulose and hemicellulose.\cite{meng2021understanding}
BSU I and II (Fig.\ref{fig:BSUs_I}, \ref{fig:BSUs_II}) have similar elemental compositions, i.e., H/C and O/C ratios, but different structures and types of functional groups. 
BSU III (Fig.\ref{fig:BSUs_III}) is polycondensed linearly with higher H/C and O/C ratios and accordingly has smaller aromatic domains and \% aromaticity. The BSU IV (Fig.\ref{fig:BSUs_IV}) is larger and highly polycondensed with lower heteroatomic ratios (H/C and O/C) in comparison to the other BSUs used in this work.

\begin{figure}[htpb!]
    \centering
    \subfloat[\textbf{BCMA}, BSUs highlighted by type.]
    {
        \label{fig:BCMA}
        \includegraphics[width=0.45\textwidth]{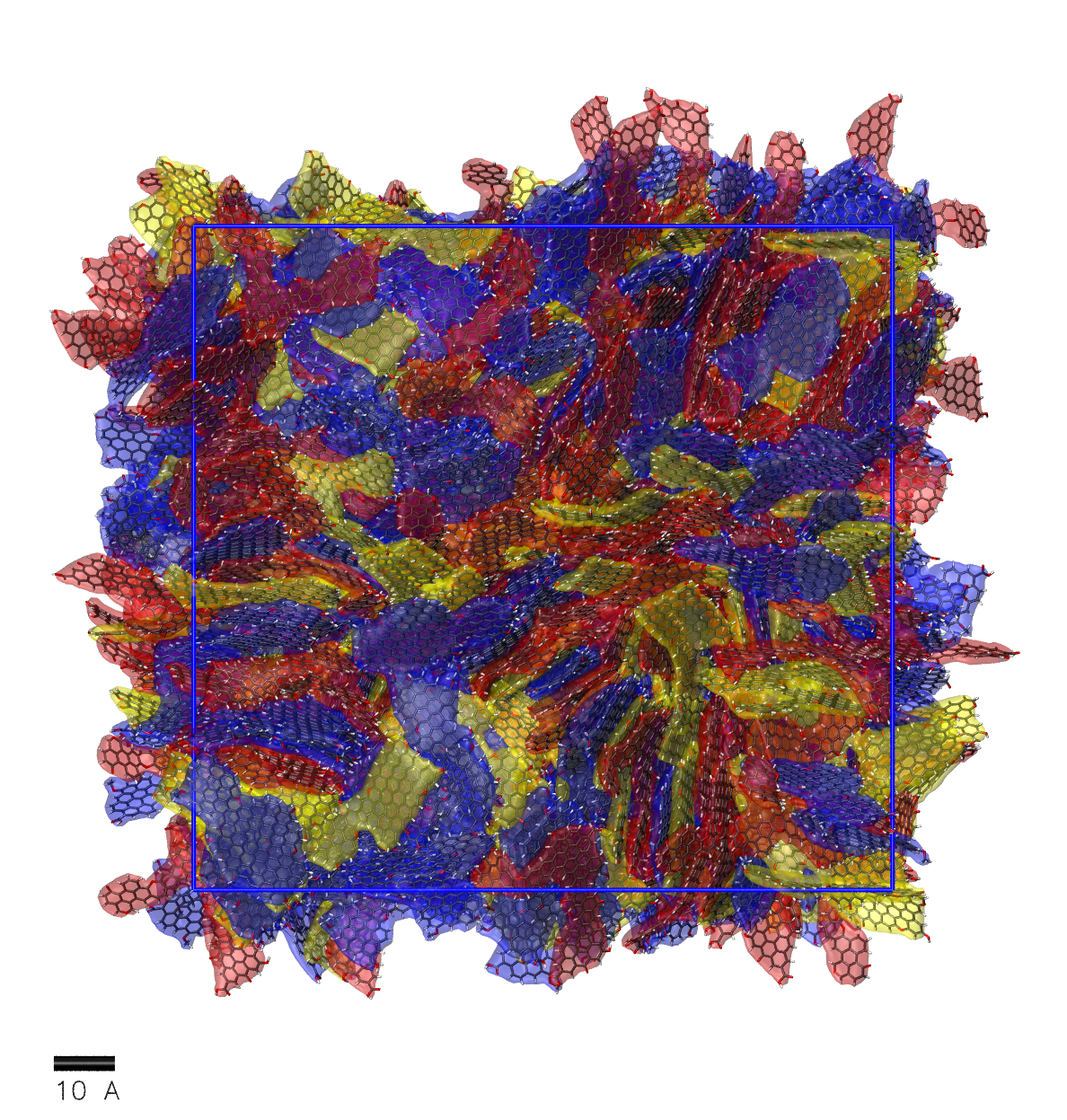}
    }\hfill
    \subfloat[\textbf{BCMB}, BSUs highlighted by type.] 
    {
        \label{fig:BCMB}
        \includegraphics[width=0.45\textwidth]{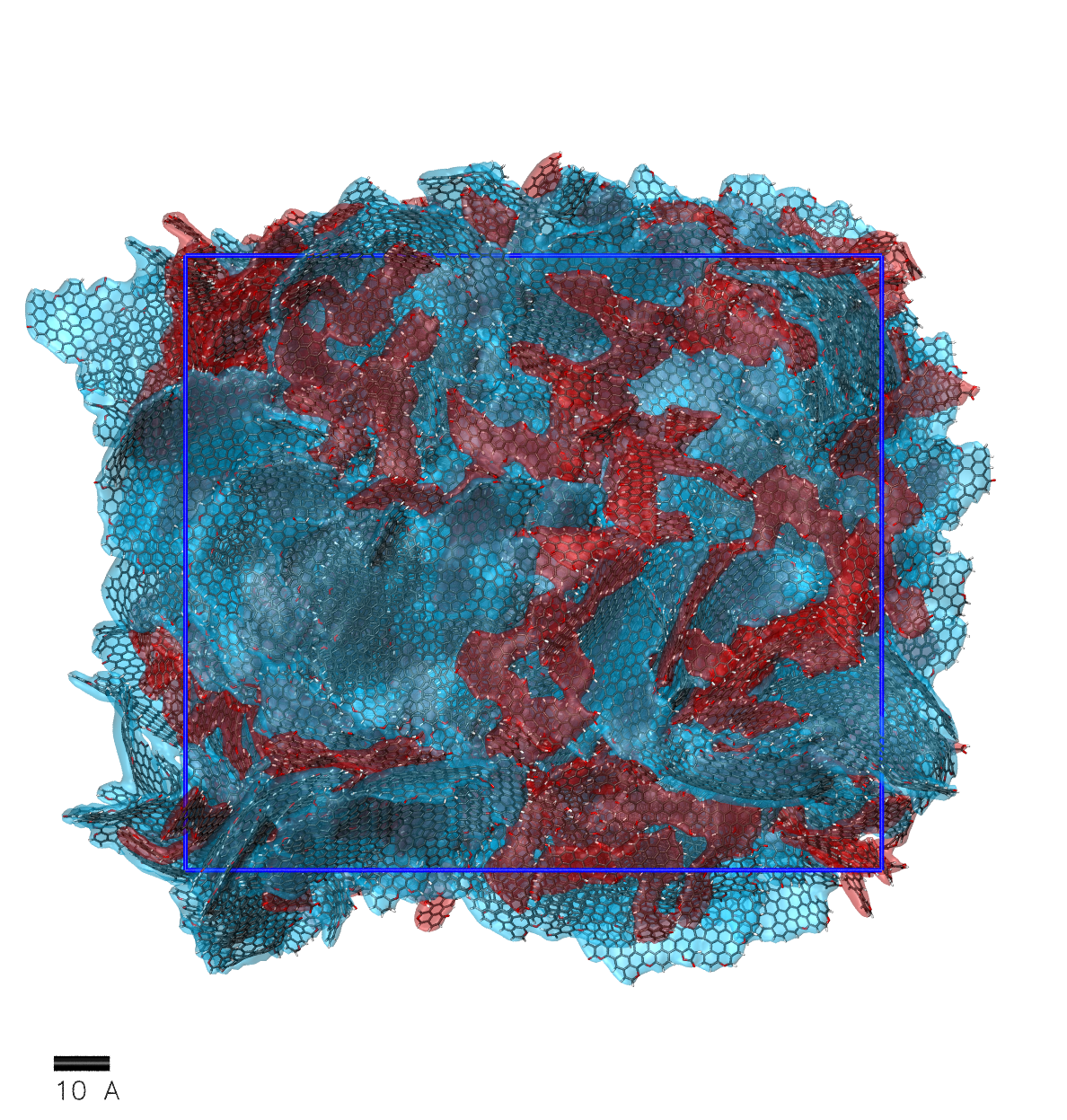}
    }\hfill
    \newline
    \subfloat[Pore size distribution and cumulative pore volume of BCMA and BCMB.] 
    {
        \label{fig:BCM_noVA_pore}
        \includegraphics[width=0.95\textwidth]{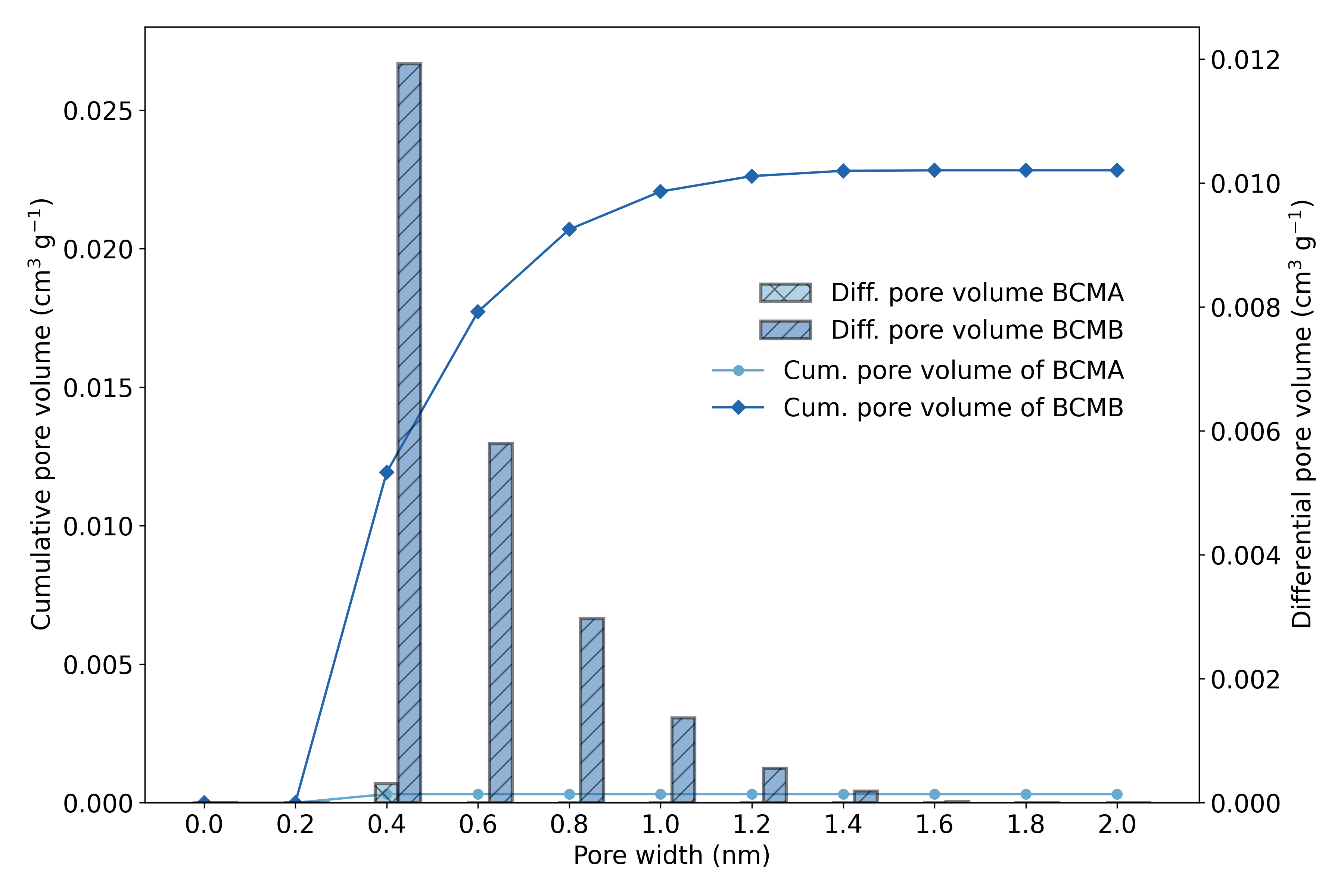}
    }\hfill
    
    \caption{Rendering of simulated biochar models BCMA (a) and BCMB (b) and the pore size distribution for these models (c). Colours for the renderings on (a) and (b): C - grey, O - red, H - white; BSUs are shown in semi-transparent, BSU I - yellow, BSU II - blue, BSU III - red and BSU IV - cyan; the periodic simulation box is shown in blue, scale bar of 1 nm is given on the bottom left.}
    \label{fig:BCM_noVA}
\end{figure}

The approach of constructing biochar from the discrete BSUs enables us to have control over the bulk properties and surface functional groups, as well as to understand the interactions between these units. 
To this end, we have set up two biochar models, BCMA and BCMB, shown in Figures \ref{fig:BCMA} and \ref{fig:BCMB}, respectively. Both models have the same average H/C (0.32) and O/C (0.08) ratios (see Table \ref{tab:BCM_noVA}).  Biochar model BCMA is composed of BSU I, II and III in the ratio 0.8:1:1. All the structural units of BCMA have similar sizes and elemental ratios. Biochar model BCMB is composed of BSU III and IV in a ratio of 3:1. As seen from Figure \ref{fig:BSUs},  these structural units vary in their, aromatic domain sizes, functional groups, elemental ratios and aromaticity.

\begin{table}[htpb!]
    \centering
    \begin{tabular}{P{4em}  P{4em} P{4em} P{4em} P{4em} P{4em} P{4em}}
    \toprule
       Biochar model  & H/C & O/C & \% Aromatic carbon  &  True density (kg m$^{-3}$) & Cum. pore volume (cm$^3$g$^{-1}$) & \% porosity \\
    \midrule
         BCMA & 0.32$\pm$0.07 & 0.08$\pm$0.02 & 67$\pm$6 & 1488$\pm$3 & 0.0001 & 0.63 \\
         BCMB & 0.32$\pm$0.14 & 0.08$\pm$0.04 & 68$\pm$13 & 1542$\pm$4 & 0.022 & 6.03  \\
    \bottomrule
     \end{tabular}
    \caption{Physiochemical properties of woody biochar models BCMA and BCMB.}
    \label{tab:BCM_noVA}
\end{table}

Table \ref{tab:BCM_noVA} presents the properties of biochar models after condensation with the simulated annealing. The true density was determined by inserting a probe molecule of Helium. The obtained true densities of 1488 kg m$^{-3}$ and 1542 kg m$^{-3}$ for biochar models BCMA and BCMB, respectively, are within the range of the target density (1546±93 kg m$^{-3}$) of the woody biochar. 
Model BCMB accommodated more intrinsic porosity (6 \%) within the structure compared to model BCMA (0.6 \%),  The biochar model BCMB had pores with sizes as large as 0.8 nm, while BCMA only featured pores up to 0.4 nm in diameter (see Figure \ref{fig:BCM_noVA_pore}). However, it must be noted that in these models, most of the pores within the generated models are in the ultra-micropore range ($<$ 0.8 nm). 
It can be seen that the choice of BSUs plays a vital role in the final morphological structure of the model. The slightly higher porosity in model BCMB is due to the large size and structure of the BSU IV. BSU IV contains numerous ring defects (5 and 7-membered rings) that bring about plane distortion, favouring the poor packing of the discrete structural units. The presence of these ring defects has previously been confirmed through HR-TEM images of biochar, in particular in the less dense regions (amorphous), in contrast to the regions of high crystallinity of stacked planar graphitic sheets.\cite{xiao2017direct}   

A higher true density of simulated biochar models correlates to higher accessible pore volumes. This corroborates with experimental findings where biochar materials with higher porosity exhibited higher true densities.\cite{brewer2014new} 
Hence, the features of the BSUs comprising biochar can affect structural properties, and the appropriate choice will make a difference in the properties, such as intrinsic porosity. 
This suggests that biochars with lower pore volumes are more likely to be composed of smaller structural units of similar sizes that pack efficiently, while those with higher pore volumes would consist of a mixture of structural units of different sizes, where the ultra-micropores arise from poor packing due to distortion caused by odd-membered rings on the large basic structural units. 
It is clear that to build biochar models with some degree of intrinsic porosity with our method, parameters such as the BSU size and the presence of ring defects (odd-membered ring) in these building units are key. This shows that bulk properties from the elemental analysis can only indicate the aromaticity index but not the structure of the basic constituents of the material. The differences brought about by the constituents of biochar play vital roles in the overall properties and functioning of the materials. 

\subsection{Biochar models with added microporosity}
\label{Ch:BCVA}

As outlined in previous sections, the original biochar models BCMA and BCMB featured predominantly ultra-micropores which are mostly accessible experimentally by CO${_2}$. Therefore, in this section, we investigate the creation of relatively larger pores by inserting virtual atoms (VAs). 
VAs studied in the simple hydrocarbon systems were used for this purpose.  We have chosen to proceed with the V10-6 ($\sigma_{V}$ = 1.0 nm, $\epsilon_{V}$ = 10$^{-6}$ kJ mol$^{-1}$) and V30-6  ($\sigma_{V}$ = 3.0 nm, $\epsilon_{V}$ = 10$^{-6}$ kJ mol$^{-1}$), as those VAs have offered a variety of the pore volumes and sizes, with a sufficient degree of softness. 

\begin{figure}[htpb!]
    \centering
    \subfloat[\textbf{BCMA\_V10} with virtual atoms V10-6.]
    {
        \label{fig:BCMA_V10}
        \includegraphics[width=0.46\textwidth]{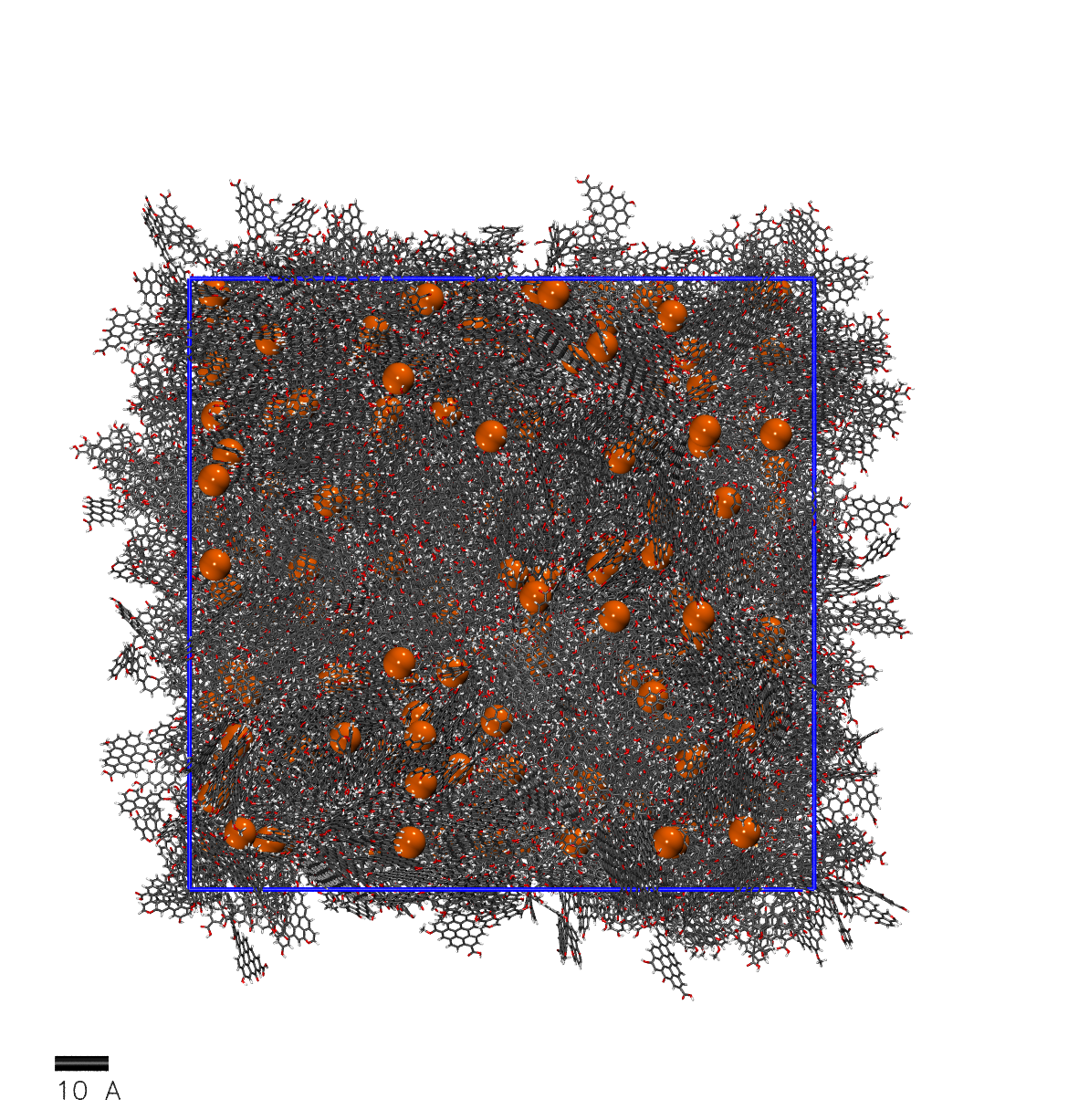}
    }\hfill
    \subfloat[\textbf{BCMA\_V30} with virtual atoms V30-6.]
    {
        \label{fig:BCMA_V30}
        \includegraphics[width=0.42\textwidth]{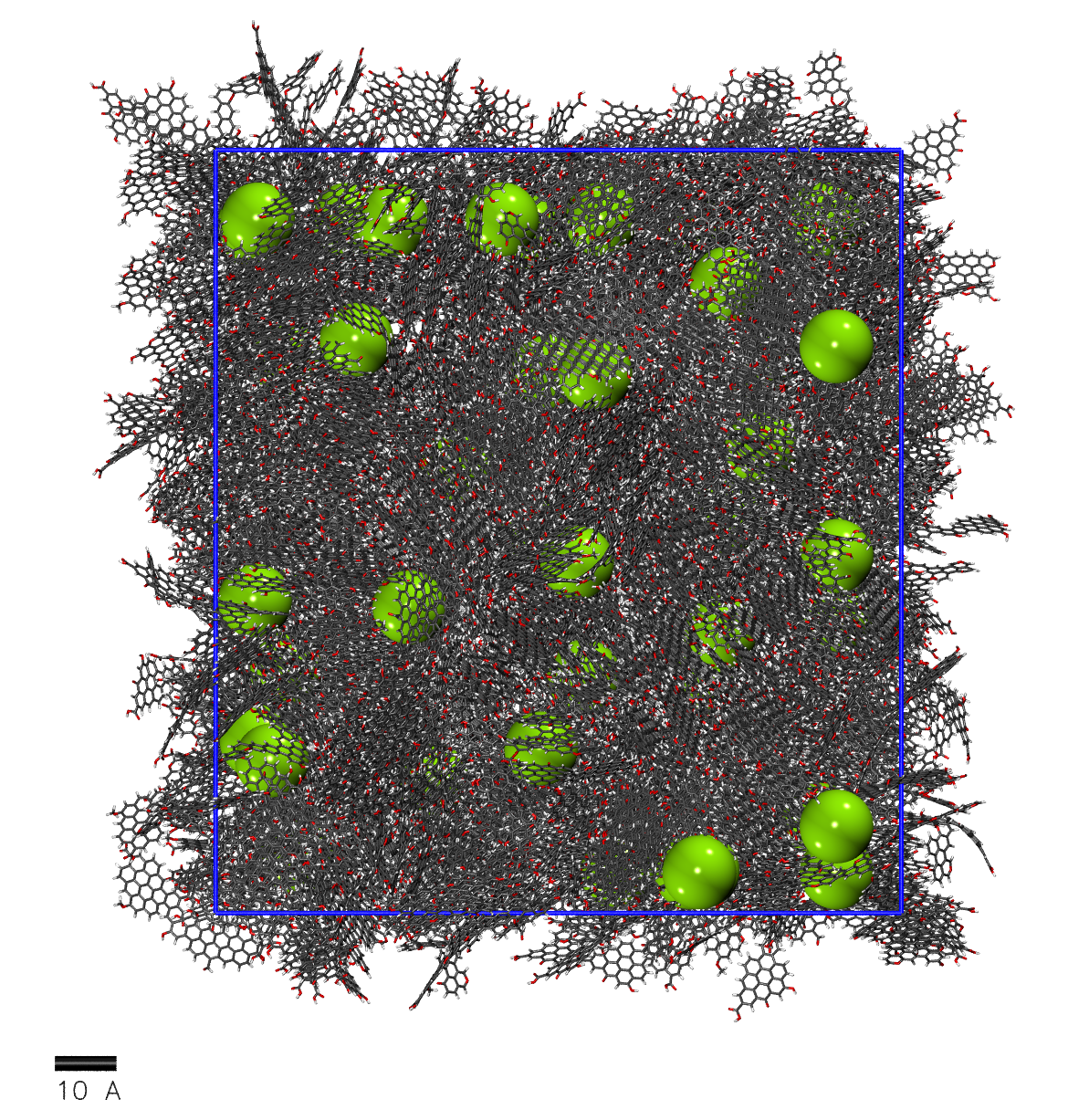}
    }\hfill
    \newline
        \subfloat[\textbf{BCMB\_V10} with virtual atoms V10-6.] 
        {
        \label{fig:BCMB_V10}
        \includegraphics[width=0.46\textwidth]{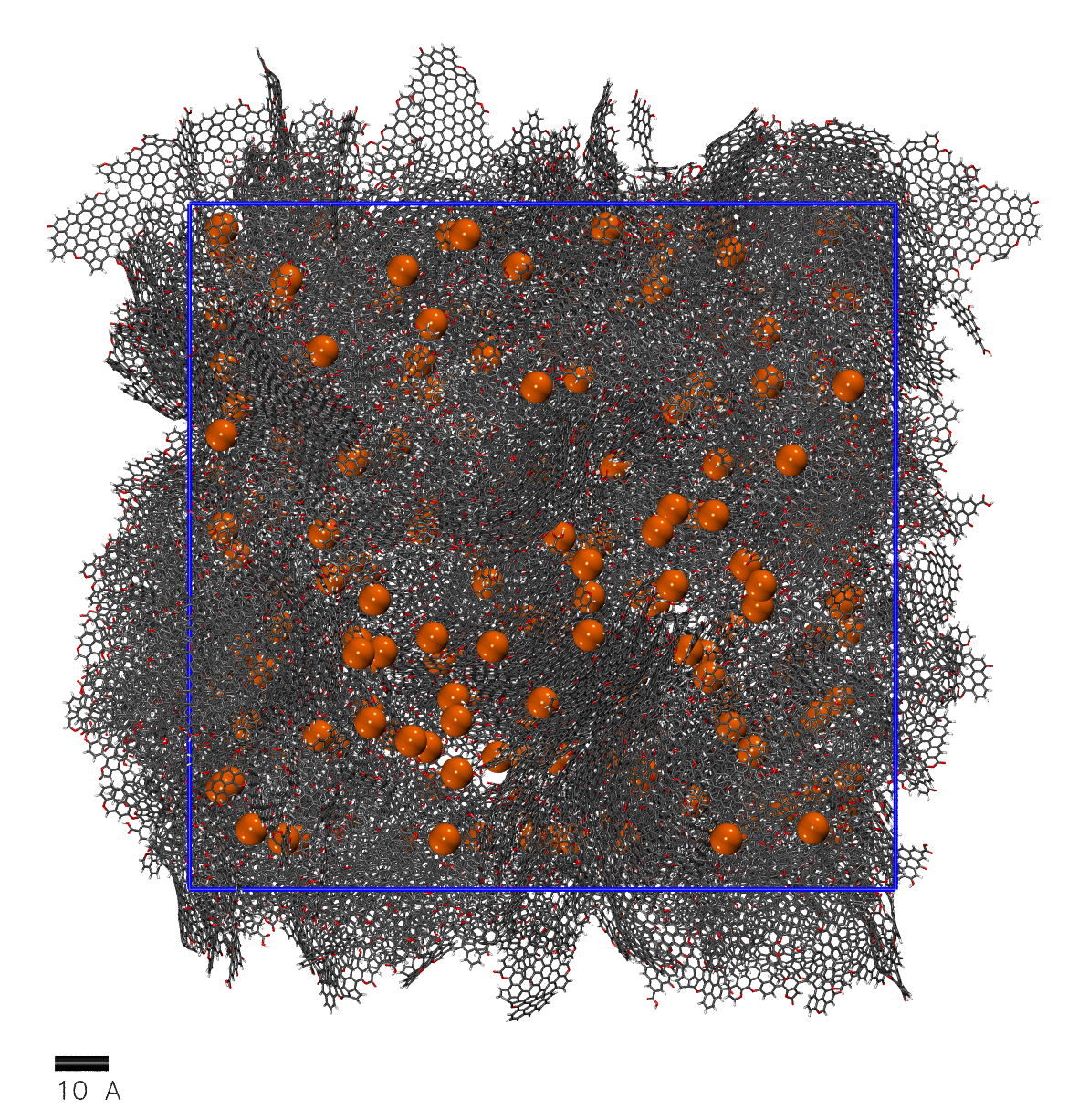}
    }\hfill
    \subfloat[\textbf{BCMB\_V30} with virtual atoms V30-6.] 
    {
        \label{fig:BCMB_V30}
    \includegraphics[width=0.46\textwidth]{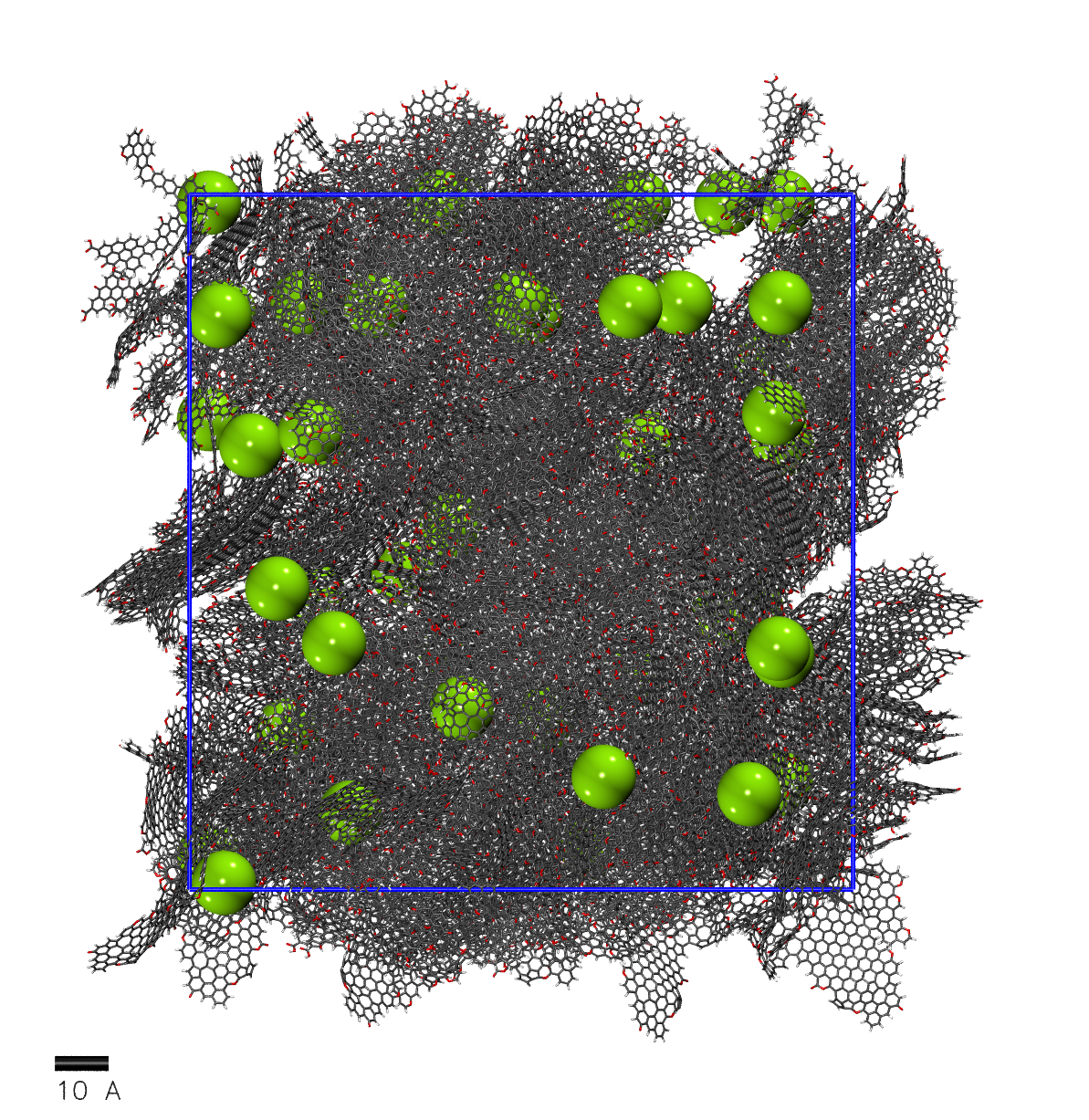}
    }\hfill
        \caption{Biochar models with virtual atoms. Colours: C - grey, O - red, H - white; Virtual atoms shown as spheres proportional to the radius, V10-6 - orange spheres, V30-6 - lime spheres; the periodic simulation box is shown in blue, scale bar of 1 nm is given on the bottom left.}
    \label{fig:BCM_VA}
\end{figure}

We have included into each of the BCMA and BCMB either 231 V10-6 virtual atoms or 49 V30-6, both produce equivalent total pore volume (cum. pore volume $\sim$ 0.1 cm${^3}$ g$^{-1}$ for the given systems), calculated according to the values presented in the Table \ref{tab:PoreVol}. Including these virtual atoms enables us to gain control over the pore sizes created in the biochar models. Biochar models with virtual atoms are labelled as BCMA\_V10, BCMB\_V10 for the models BCMA and BCMB, respectively, with the virtual atom V10-6 and BCMA\_V30 and BCMA\_V30 for the models with V30-6 virtual atoms. The condensed materials are shown in Figure \ref{fig:BCM_VA}.

As expected, biochar models developed with virtual atoms had higher total pore volumes with a wider distribution of pore sizes compared to the biochar models generated without virtual atoms. The generated pore sizes increased from a maximum of 0.4 nm in BCMA to 0.8 nm (BCMA\_V10) and 1.2 nm (BCMA\_V30).  For the BCMB model, the increase was from 1.4 nm without virtual atoms to 2.2 nm (BCMB\_V10) and 2.4 nm (BCMB\_V30), respectively.
The pore sizes of interest in this work are in the micropore size ranges and have been reported in the literature from gas adsorption studies\cite{sigmund2017biochar,maziarka2021tailoring}. These micropores contribute massively to the surface area of biochar materials, which is a desirable property in particular for gas filtration. These pores also play a role in defining the structure of the material at the nanoscale. From our developed biochar models we gain control over the pore size distribution, implying that one could generate biochar models of larger sizes and wider distribution of pores by mixing virtual atoms of different sizes. 

The total pore volume of the biochar models is a result of the cumulative sum of all pores. In the simulated biochar model the total pore volume of BCMA increased from 0.0001 cm$^3$ g$^{-1}$ to 0.003 cm$^3$ g$^{-1}$( BCMA\_V10) and 0.008 cm$^3$ g$^{-1}$ (BCMA\_V30). Similarly, the total pore volume in the biochar model BCMB increased from 0.042 cm$^3$ g$^{-1}$ to 0.185 cm$^3$ g$^{-1}$(BCMB\_V10) and 0.138 cm$^3$ g$^{-1}$(BCMB\_V30). The higher pore sizes and volumes in models BCMB result from the poor packing of the BSUs IV and III, which is further distorted by the VAs. 

However, upon removal and re-equilibration of the systems, the pore volume of the biochar models collapsed to various extents. The pore volumes in BCMA\_V10 and BCMA\_V30 dropped to a greater degree compared to their BCMB counterparts. For both BCMA\_V10 and BCMA\_V30, the total pore volume decreased by about 33\% and 25\%, respectively, while the pore total volume dropped by 15\% and 8\% in BCMB\_V10 and BCMB\_V30, respectively. The greater collapse of pore volume in BCMA models is linked to the structure and sizes of the BCMA model building blocks. The BSUs used are relatively smaller, with few ring defects and, as such, flat and planar, which favours close-packed arrangements. Biochar models BCMB, on the other hand, contain numerous odd-membered rings which cause distortion away from planarity forming curvatures that can better support and maintain the pores created by the virtual atoms even after those were removed. The changes in the pore size and volumes in the generated biochar models before and after the removal of the virtual atoms are shown in SI Fig S3. The final cumulative pore volumes for each system are given in Table \ref{tab:BCM_poresize}.

The cumulative micropore (pore width $\leq$ 2 nm) volumes of 0.118 cm$^3$ g$^{-1}$ for woody biochar produced at 550 °C, and volumes of 0.175 and 0.135 cm$^3$ g$^{-1}$ for biochars produced at 700 °C and 500 °C, respectively, are reported in the literature.\cite{sigmund2017biochar, maziarka2021tailoring} Therefore, our models BCMB\_V10, with the cumalative pore volume of 0.157 cm$^3$ g$^{-1}$, and BCMB\_V30, with the volume of 0.127 cm$^3$ g$^{-1}$, are in a good agreement with the experimental measurements. 

\begin{table}[htpb!]
    \centering
    \begin{tabular}{ccc}
        \hline
         Model&  Cum. pore volume (cm$^3$ g$^{-1}$)& \% porosity \\
         \hline
    BCMA\_V10 &  0.002 & 1.61  \\
    BCMA\_V30 &  0.006 & 2.92 \\
    BCMB\_V10 & 0.157 & 25.4 \\
    BCMB\_V30 & 0.127 & 20.7 \\
    \hline
    \end{tabular}
    \caption{Cumulative pore volume and \% porosity for the BCMA and BCMB biochar models, set up with virtual atoms V10-6 and V30-6, which were then removed.}
    \label{tab:BCM_poresize}
\end{table}

\begin{figure}[htpb!]
    \centering
    \subfloat[Pore volume analysis for the model BCMA produced with virtual atoms.]
    {
        \label{fig:BCMA_PoreVol}
        \includegraphics[width=0.9\textwidth]{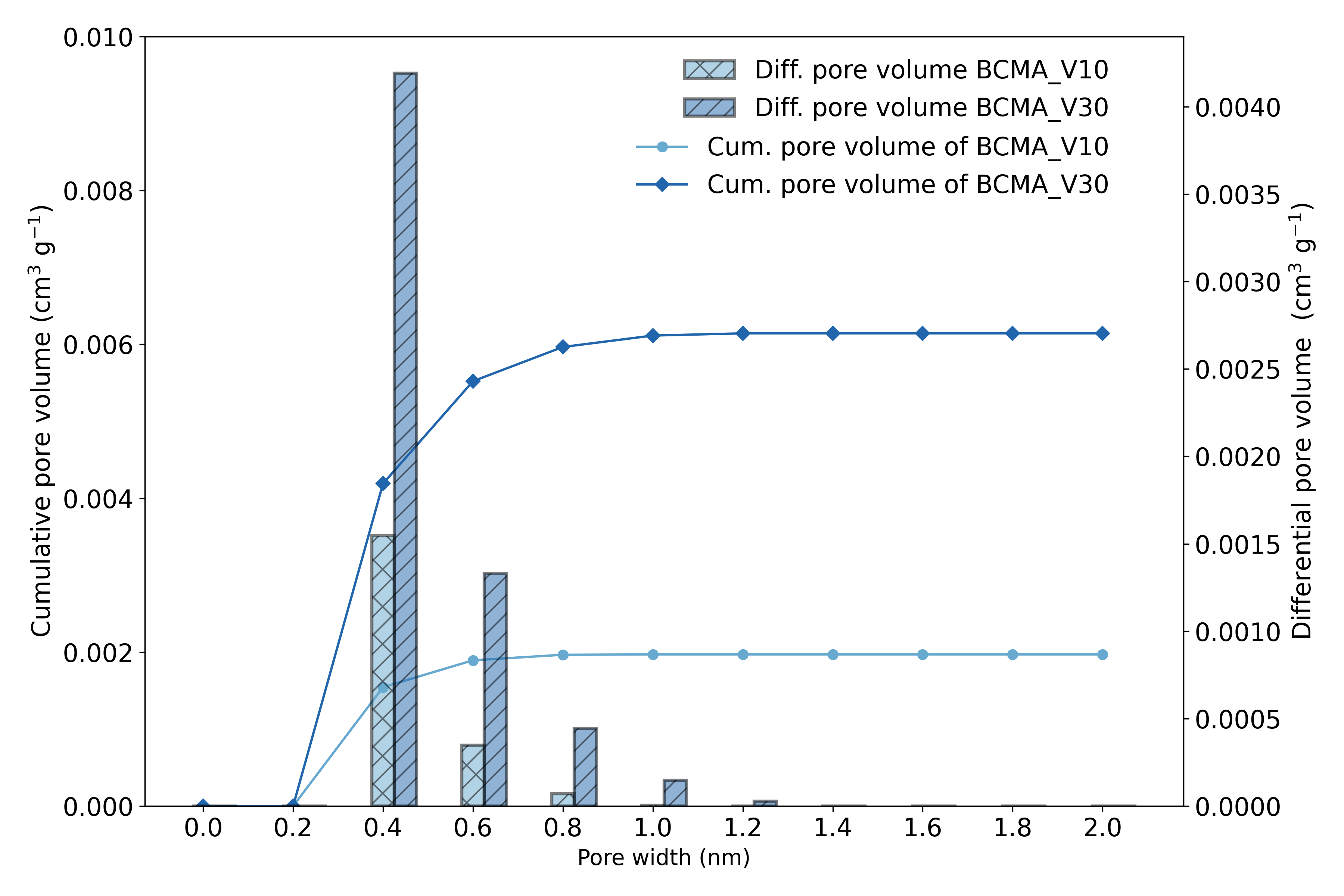}
    }\hfill
      \newline
        \subfloat[Pore volume analysis for the model BCMB produced with virtual atoms.] 
        {
        \label{fig:BCMB_PoreVol}
        \includegraphics[width=0.9\textwidth]{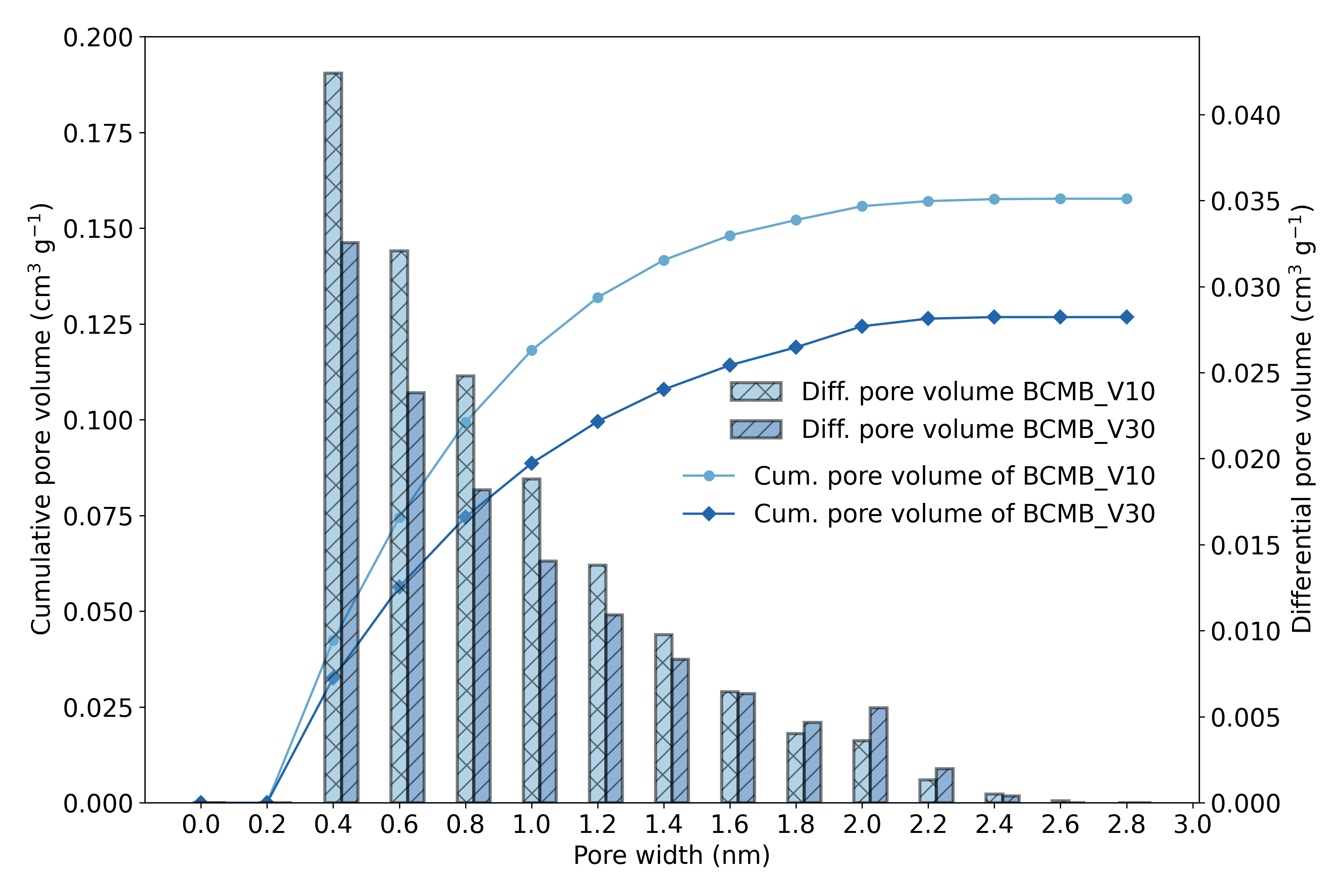}
    }\hfill
    
        \caption{Cumulative pore volume and differential pore volume of (a) BCMA and (b) BCMB biochar models after the removal of virtual atoms V10-6 and V30-6.}
    \label{fig:BCM_PoreVol_plot}
\end{figure}

\begin{figure}[htpb!]
    \centering
    \subfloat[BCMA.]
    {
        \label{fig:BCMA_sasa}
        \includegraphics[width=0.3\textwidth]{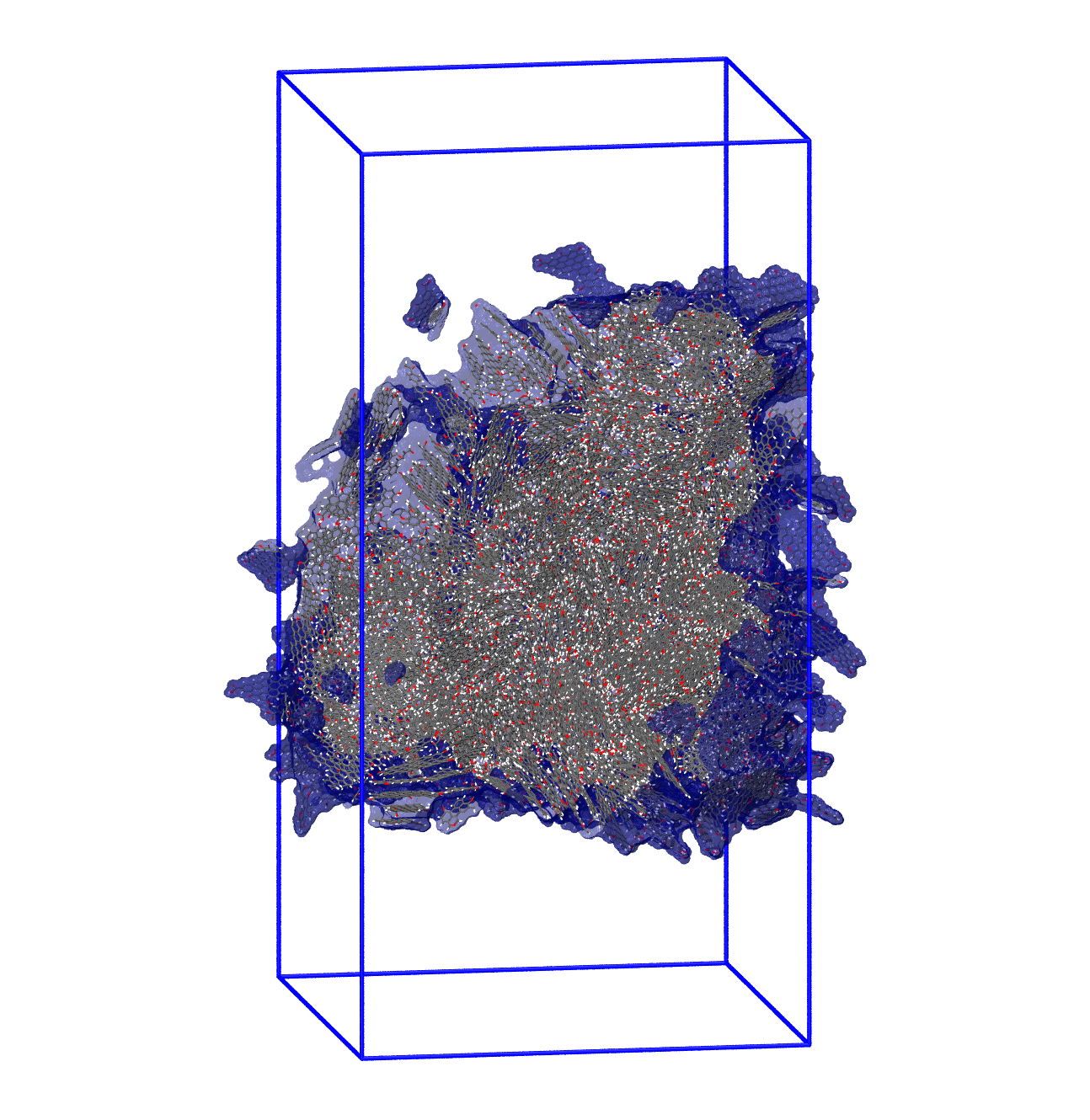}
    }\hfill
    \subfloat[BCMA\_V10.] 
    {
        \label{fig:BCMA_V10_sl}
        \includegraphics[width=0.3\textwidth]{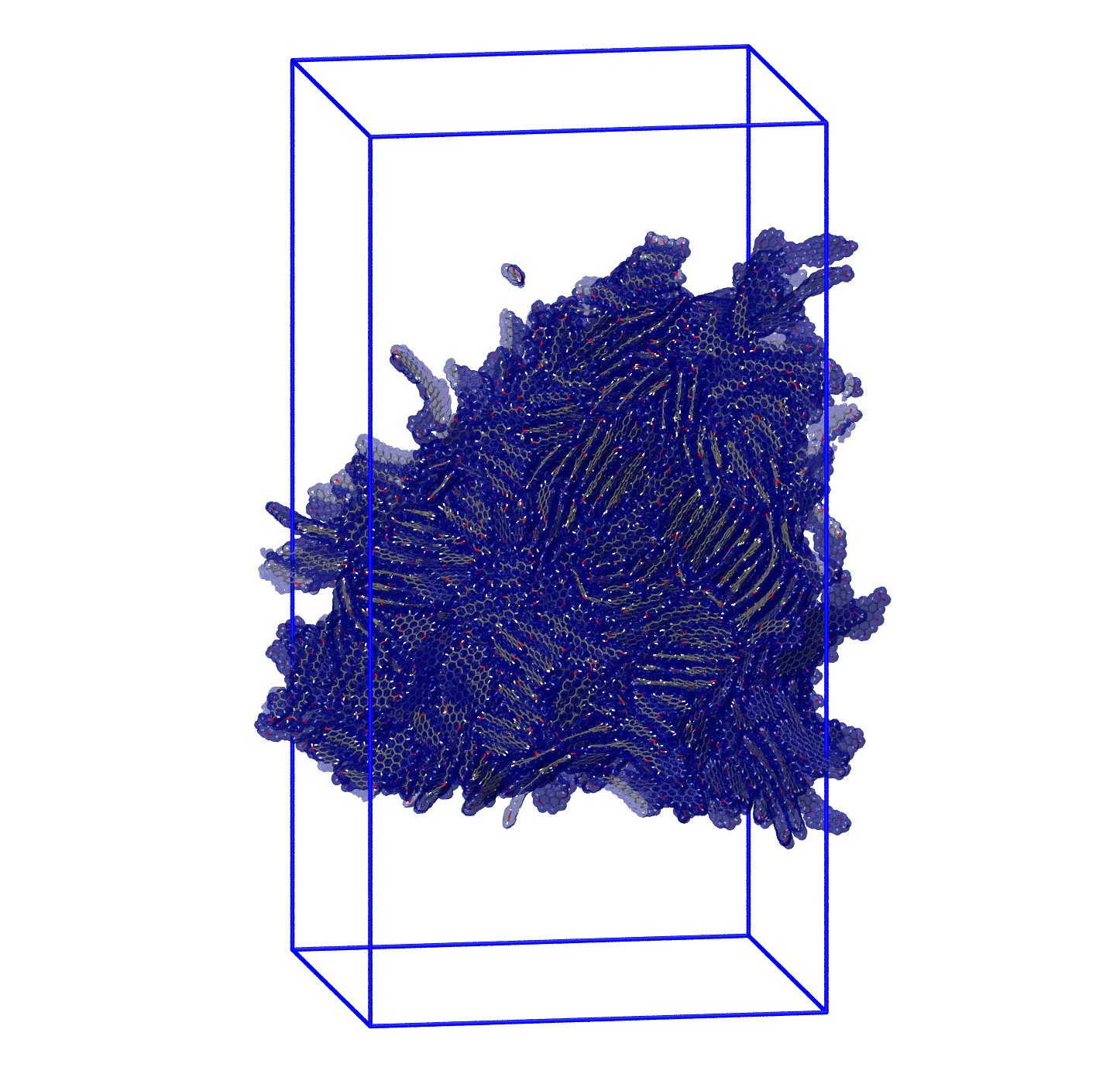}
    }\hfill
    \subfloat[BCMA\_V30.] 
        {
        \label{fig:BCMA_V30_sl}
        \includegraphics[width=0.3\textwidth]{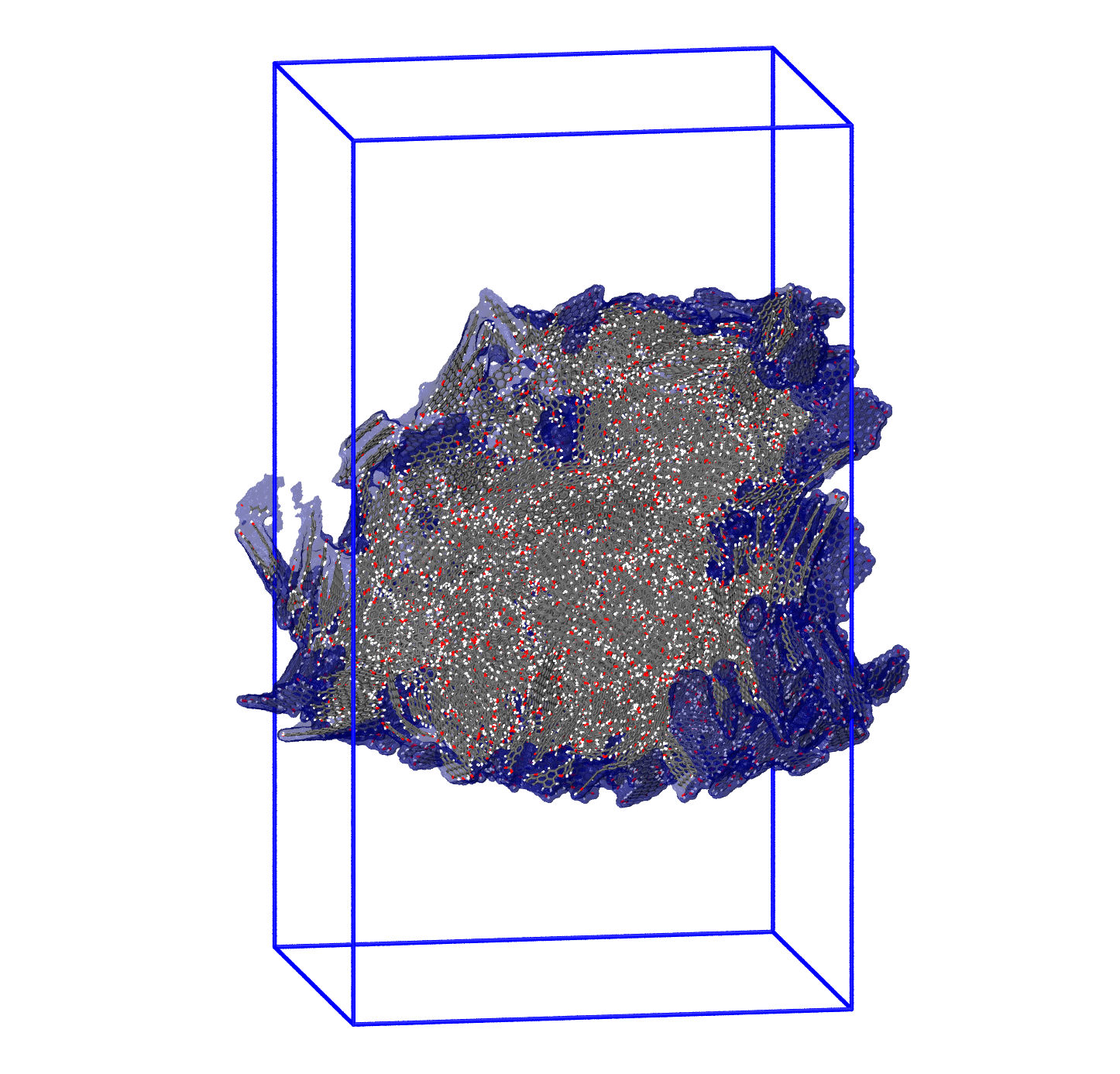}
    }\hfill
   \newline
    \subfloat[BCMB.]
    {
        \label{fig:BCMB_sasa}
        \includegraphics[width=0.3\textwidth]{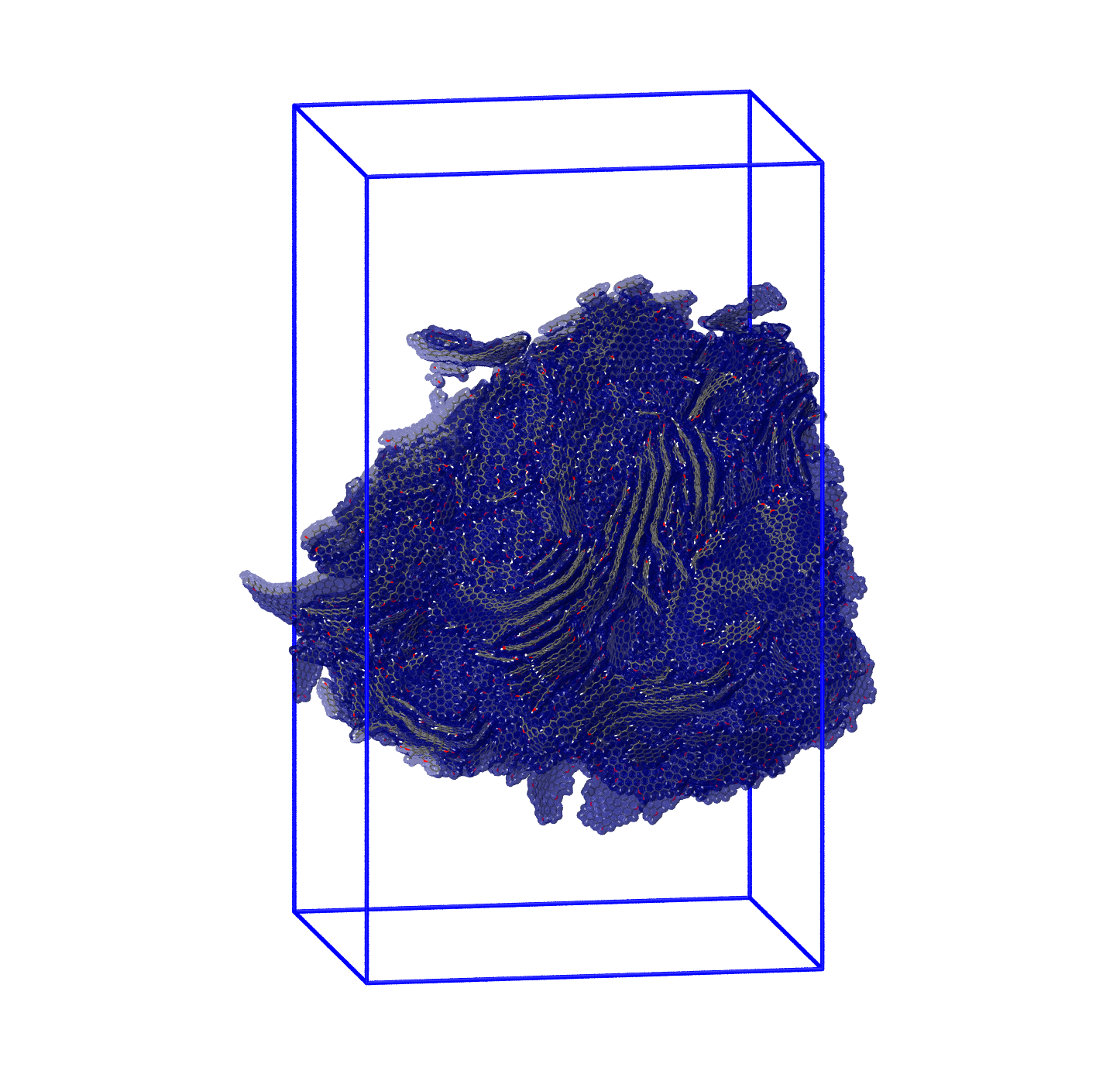}
    }\hfill 
    \subfloat[BCMB\_V10.] 
    {
        \label{fig:BCMB_V10_sl}
    \includegraphics[width=0.3\textwidth]{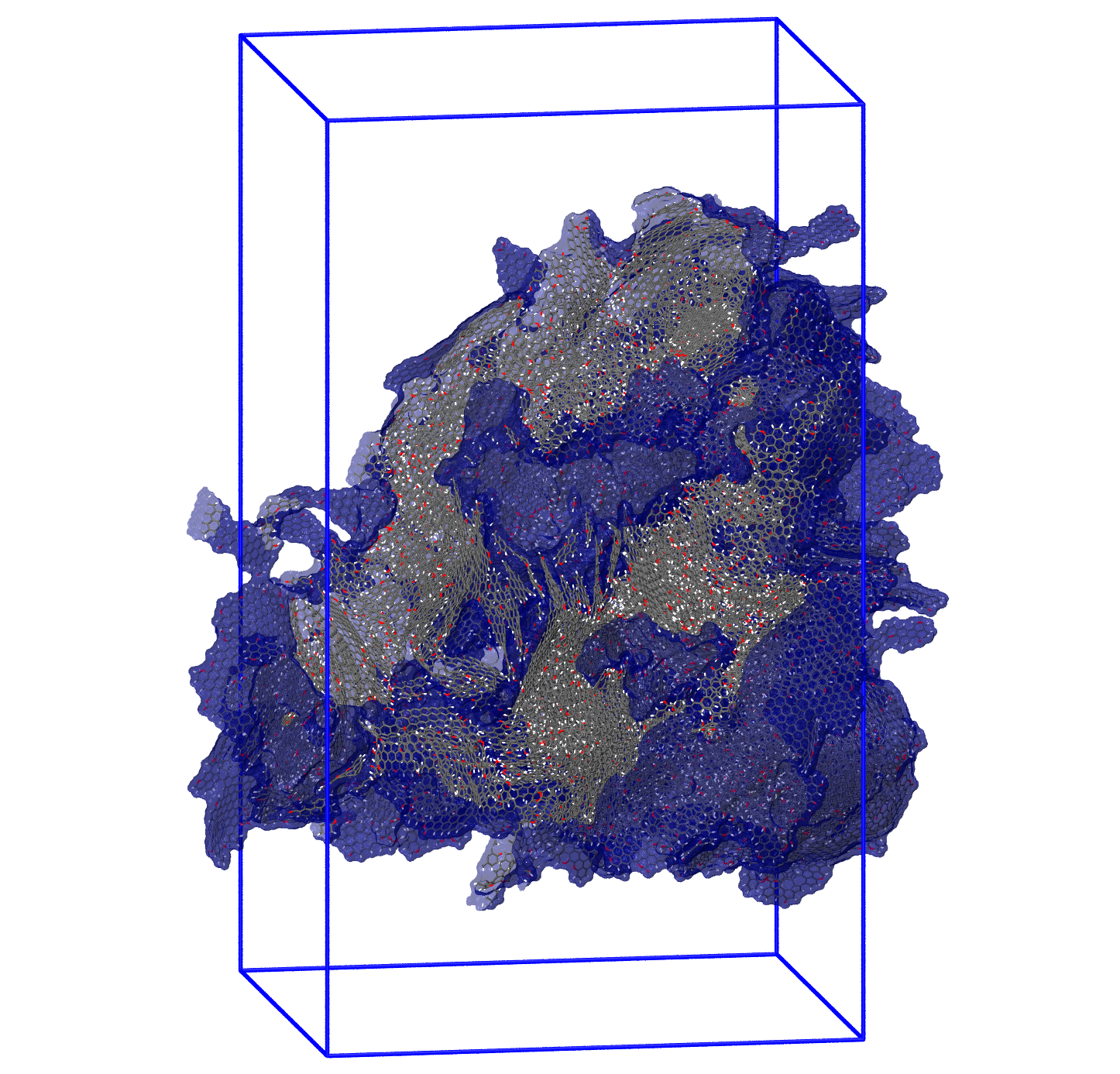}
    }\hfill
    \subfloat[BCMB\_V30.] 
    {
        \label{fig:BCMB_V30_sl}
    \includegraphics[width=0.3\textwidth]{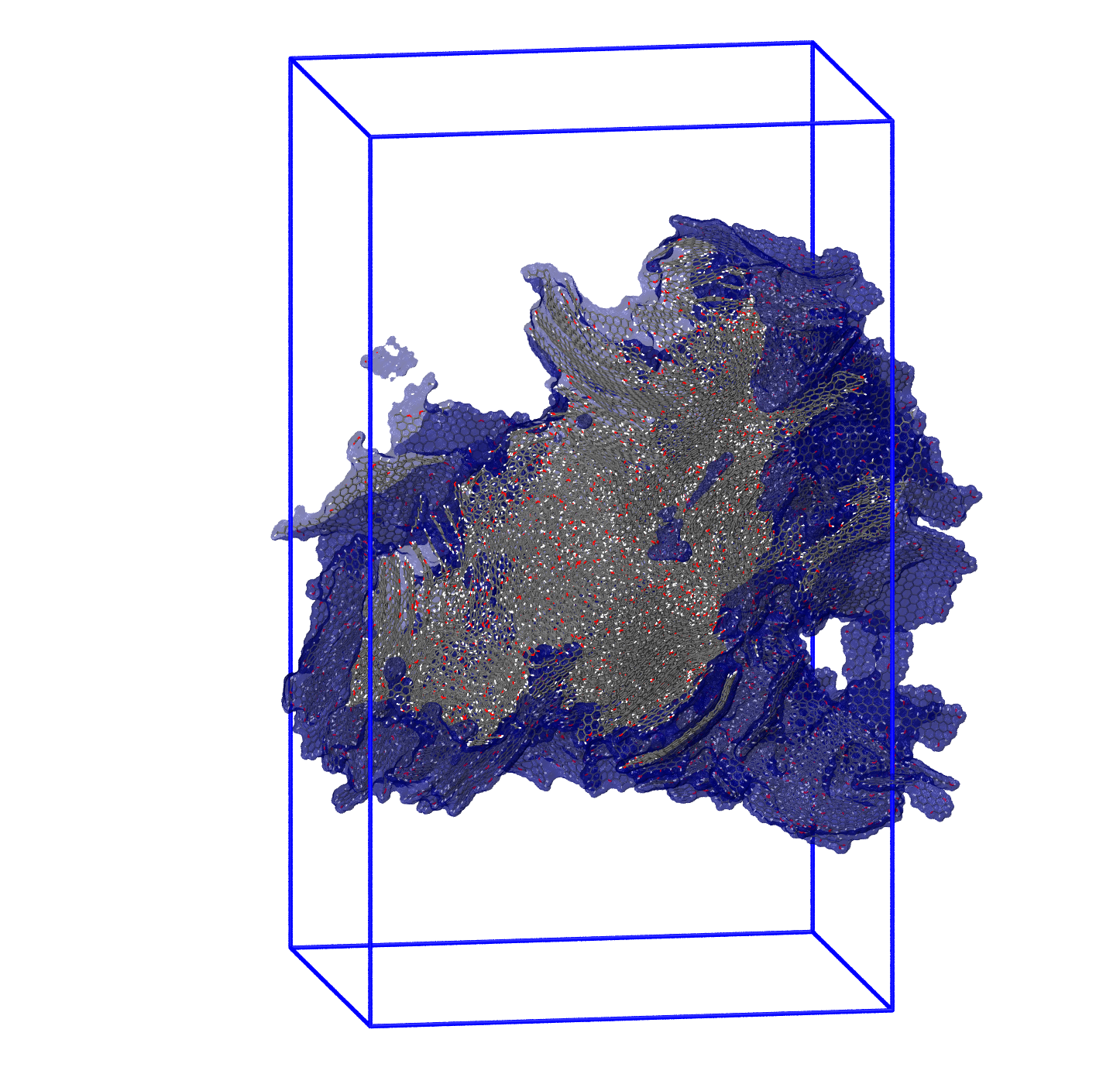}
    }\hfill
        \caption{Porous biochar models with exposed surfaces in z-direction. Colours: C - grey, O - red, H - white.  The periodic simulation box is shown in blue. SASA (calculated with N$_2$ probe)  is shown as blue surface. The renderings show a slice through each model along (111) plane to allow visualisation of internal structure and highlight the accessible pores. }
    \label{fig:BCM_surface}
\end{figure}

Biochar pores span a wide range of sizes, while micropores have been reported to contribute the most to the surface area of the material \cite{weber2018properties,lehmann2007biochar}. The surface area is a key desirable property as it defines the adsorption capacity and drives the interactions with the surrounding species.
While micropores contribute to surface area, not all pores within a given biochar are accessible. Condensing tar could clog pores, making them inaccessible, as well as some pores will remain filled with the volatile compounds, which never escaped, but instead condensed upon cooling within the pore. 
Therefore, we measure the solvent-accessible surface area (SASA) of the surface-exposed biochar models. Since SASA is typically reported per weight, which is not meaningful in the context of molecular models, where more or less bulk can be added into the simulation cell, we also report normalised SASA per xy-cross-sectional area of the simulation box, Table \ref{tab:4}. The SASA and normalised SASA of BCMA models were two times lower than BCMB models. This again highlights the importance of BSUs used in the biochar models in defining the packing and surface roughness of the biochar model. The normalised SASA is linearly related to the pore volumes of the simulated models. These findings are in good agreement with the trends observed from experiments. The renderings of the surface exposed biochar models are shown in Fig \ref{fig:BCM_surface}. 

\begin{table}[htpb!]
    \centering
    \begin{tabular}{c c c }
    \toprule
       Biochar model & Average SASA (m$^2$ g$^{-1}$) & Average normalised SASA  \\
    \midrule
         BCMA & 238.75 & 1.98\\
         BCMA\_V10 & 257.57 & 2.12\\
         BCMA\_V30 & 283.11 & 2.30\\
         BCMB & 402.73 & 3.36\\
         BCMB\_V10 & 691.08 & 5.00\\
         BCMB\_V30 &  543.14 & 4.11\\
    \bottomrule
    \end{tabular}
    \caption{Surface accessible surface area (SASA) of biochar models.}
    \label{tab:4}
\end{table}

\section{Conclusions}

In this work we present an iterative molecular dynamic approach to generate biochar models with controlled porosity. The approach combines different basic structural units, here a selection of molecular building blocks and virtual atoms that create pore space between the molecular units. We demonstrate the approach on the development of biochar models, representative of woody biochar produced at 600 -- 650 °C. Our models reproduce the chemical descriptors (H/C, O/C atomic ratios, \% aromatic carbon, aromatic domain sizes)  and physical properties (true density, cumulative porosity and pore size distribution). 

With the iterative approach, we gain control over the composition of biochar models and are able to identify key parameters to consider when developing realistic biochar models. In particular, the choice of the structure and size of molecular building blocks influences the final textural and morphological properties of obtained biochar models. Using building blocks with aromatic domain sizes $\leq$ 33 resulted in mostly nonporous structures. While integration of building blocks with large aromatic domain sizes  ($>$400), featuring non-hexagonal rings, allows for the development of micropores within the biochar matrix.
The addition of virtual atoms, which are soft repulsive Lennard-Jones spheres, during the condensation step allows for integration of desired porosity in the produced biochar models.
Microporous biochar models with pore sizes in the range of 0.4 -- 2.6 nm were obtained with virtual atoms with $\sigma = 1.0$ nm and $3.0$ nm (both with $\epsilon = 10^{-6}$ kJ mol $^{-1}$), suggesting that by mixing virtual atoms of different sizes models with a wider pore size distribution can be produced. 
In this work, the cumulative micropore volume (pores widths $\leq$ 2 nm) for the developed models BCMB\_V10 (0.157  cm$^3$ g$^{-1}$) and BCMB\_V30 (0.127 cm$^3$ g$^{-1}$ ) were in an excellent agreement with the experimental values of 0.118 -- 0.175  cm$^3$ g$^{-1}$.  

The produced microporous biochar models are of relevance for the study of chemical interaction in confined pores, adsorption and separation of gases and smaller molecules. Combining molecular building blocks and virtual atoms, as demonstrated in this work, allows for tuneing the developed models to obtain desired properties. The biochar molecular models presented in this work, along with the building blocks and virtual atoms used to construct them, are freely available from our GitHub page:\url{https://github.com/Erastova-group/Porous_Biochars_Models}.  We hope that our shared models will remove the major hurdle in the uptake of molecular modelling by the biochar research community, while our approach will assist in the development of knowledge of the biochar properties and functionality at the nanoscale. We believe that the combined use of molecular simulations and experimental characterisation is the key to enable knowledge-driven design and development of new products in an efficient and sustainable way.

\section{Methodology}
\label{Ch:Methods}

\subsection{Molecular Structures and Force Fields}
\label{Ch:Method.1}

 Structures of simple hydrocarbon compounds (nonane, nonanoic acid, pentadecane, toluene, phenol and coronene) and biochar basic structural units (BSU) were drawn with the aid of MarvinSketch 21.17 software (by ChemAxon, http://www.chemaxon.com).  All simulations were performed using OPLS all-atom force field.\cite{jorgensen1996development} The force field parameters for the biochar BSU were assigned using PolyParGen\cite{yabe2019development}.

Simple hydrocarbon systems with $\sim$10000 atoms were prepared each in a 10 x 10 x 10 nm simulation box. Biochar models were prepared with $\sim$ 126000 atoms in a simulation box of 20 x 20 x 20 nm in size. 

Virtual atoms (VAs) were used to create pores. In pure hydrocarbon systems, a maximum of 5 VAs were randomly distributed in the systems at distances greater than the \text{$\sigma_{V}$} of the virtual atom used. The parameters of the pores created by the VAs in the simple hydrocarbon systems were used to determine the required number of VAs for the target pore volumes in biochars. The VAs were randomly distributed at distances greater than the size of the VA used.

\subsection{Simulation details}
\label{Ch:Sim_method}

Molecular dynamics simulations were performed using open-source software GROMACS 2022.3.\cite{abraham2015gromacs}. 
The simulations for all systems were first energy minimized using the steepest descent algorithm with the convergence criterion where the maximum force on any one atom is less than 500 kJ mol$^{-1}$ nm$^{-1}$.  

Subsequently, equilibration simulations were carried out in the isothermal-isobaric ensemble (NPT) ensemble with a time step of 1 fs, cut-off of 1.5 nm and periodic boundary conditions. The simulations were run with a real-space particle-mesh-Ewald (PME) algorithm to calculate the van der Waals force and long-range electrostatic interactions; long-range dispersion correction was applied. 
The temperature of the system was coupled using a velocity-rescaling thermostat with a time constant of 0.1 ps. The system pressure was coupled with Nosé-Hoover barostat at a time constant of 10 ps anisotropically unless otherwise mentioned. Specific details are given in the sections below.

\subsubsection{Simple hydrocarbons}
\label{Ch:simple_HC}

After energy minimisation, the equilibration was carried out as follows (summarised in Table \ref{Tab:1}):
\begin{enumerate}
    \item the systems were compressed for 3.5 ns the isotropic NPT ensemble at 100 bar and 300 K or the temperature above the melting temperature;
    \item this was followed by the simulated annealing, as the systems were cooled to 300 K at 1 bar;
    \item   for systems containing virtual atoms, after the second step, the virtual atoms were removed and the system relaxed for 10 ns at 1 bar, 300 K; 
    \item The final 20 ns simulation was performed at 1 bar, 300 K. 

The systems were assumed to be equilibrated when the density remained constant. The last 10 ns of the simulation were used for analysis.
\end{enumerate}

\newcolumntype{P}[1]{>{\centering\arraybackslash}p{#1}}

\begin{table}[h!]
    \centering    
    \begin{tabular}{P{7em}  P{6em}  P{3em} P{3em} P{4em}  P{5em} P{3em} P{4em}} 
    \hline 
         Compound &  No. mol (No. atoms) & \multicolumn{3}{c}{1$^{st}$ NPT simulation}& \multicolumn{3}{c}{2$^{nd}$ NPT simulation}\\
         \hline 
         & &T (K) & P (bar) &Sim. time (ns)  & T (K) & P (bar) &Sim. time (ns)\\ 
         \hline
         Nonane & 345 (9315) & 300 & 100 & 3.5  & 300 & 1 & 10\\ 
         Toluene & 667 (10005) & 300 & 100 & 3.5  & 300 & 1 & 10\\ 
         Phenol & 770 (10010) & 320 & 100 & 3.5  & 320 $\rightarrow$ 300 & 1 & 10\\ 
         Pentadecane & 270 (9990) & 350 & 100 & 3.5  & 350 $\rightarrow$ 300 & 1 & 24\\ 
         Coronene & 278 (10008) & 1000 & 100 &3.5  & 1000 $\rightarrow$ 700 $\rightarrow$ 300 & 1 & 39\\ 
         Nonanoic acid &345 (9315) & 300 & 100 &3.5  & 300 & 1 & 10\\ 
           \hline
    \end{tabular}
      \caption{Simulation parameters of simple hydrocarbon systems.}
    \label{Tab:1}
\end{table}

\subsubsection{Biochars}

\label{Ch:BC}
The biochar basic structural units (BSU) were assembled and energy minimized, as described above, before equilibration. 
The equilibration was carried out as follows: 
\begin{enumerate}
    \item the systems were simulated for 3.5 ns  at 1000 K and 200 bar with isotropic pressure coupling applied;
    \item simulated annealing was performed at 200 bar cooling to 300 K to build the bulk-condensed biochar molecular model. During this step, the system was annealed at 1000 K for 5 ns, then cooled to 700 K at a rate of 12 K per ns, the system was then maintained at 700 K for 10 ns, before cooling to 500 K at a rate of 25 K per ns, then the system was maintained at 500 K for 10 ns, then cooled to 300 K at a rate of 40 K per ns and maintained at 300 K for 10 ns. This resulted in a total of 63 ns simulation time;
    \item a final simulation at 1 bar, 300 K for 30 ns to enable the bulk biochar condensed system to relax at ambient conditions;
    \item For systems containing VAs, those were removed, and a further 30 ns simulation at 1 bar, 300 K was performed.
\end{enumerate}

The system was assumed to be equilibrated, when no changes in density, volume, dimensions per axis and root mean square deviation of BSUs were observed. Therefore, the analysis was carried out over the final 10 ns of the trajectory.

\subsubsection{Density, porosity and pore size distribution}

The density of the bulk condensed model represents the \textbf{bulk density}. The \textbf{true density} of simulated models is calculated from the average porosity according to eq (\ref{eqn:2}).

\begin{equation}
\rho_{\text{true}} = \dfrac{\rho_{\text{bulk}}} {1 - \phi} 
\label{eqn:2}
\end{equation} 
where; $\rho_{\text{true}}$ and $\rho_{\text{bulk}}$ are the true and bulk densities, respectively, and $\phi$ is the porosity of the bulk model. 

The \textbf{average porosities} of simulated models were assessed through the probe molecule (here we use helium with a radius of 0.13 nm) insertion technique. 

The\textbf{ pore size distribution} is evaluated by the geometric method through sphere insertion. Here, spheres of different radii are inserted within the solid biochar matrix, and the volume occupied by the inserted sphere is measured. The radius of the fitted sphere represents the radius of the pore space. 

The \textbf{solvent-accessible surface areas} (SASA) were evaluated for the exposed surfaces of the simulated biochar models using the probe radius of 0.186 nm (representative of N$_2$ gas). The SASA was normalized for each of the exposed surfaces due to the variations in the dimensions of the simulated models. The normalised SASA of the models is computed from eq (\ref{eqn:3}).

\begin{equation}
\text{Normalised SASA} = \sum_{ij}\frac{\text{SASA}}{\text{(2 x A$_{ij}$)}}
\label{eqn:3}
\end{equation}
where; A$_{ij}$ is the exposed cross-sectional area and ${ij}$ is the cross-sectional surface (xy, yz and zx).

The analysis is performed using tools within the GROMACS package\cite{abraham2015gromacs}, and further analysis and plots are carried out with in-house Python\cite{10.5555/1593511} scripts.
\subsubsection {Visualisations}
The renderings are made using VMD 1.94 \cite{humphrey1996vmd}, molecular units are shown in licorice, C - grey, O - red, H - white, BSUs are highlighted as QuickSurf in transparent, BSU I - yellow, BSU II - blue, BSU III - red, BSU IV - cyan. SASA presented on the renderings is produced by MoloVol 1.1.1 \cite{maglic2022molovol}, using a probe of 0.186 nm (N$2$ spherical radius), grid resolution of 0.02 nm and optimisation depth of 4 cycles. Plots are produced with Python 3.11,\cite{10.5555/1593511} using Matplotlib 3.8.\cite{Hunter:2007}


\begin{acknowledgement}

This work used the Cirrus UK National Tier-2 HPC Service at EPCC (www.cirrus.ac.uk) funded by the University of Edinburgh and EPSRC (EP/P020267/1).

\end{acknowledgement}

\begin{suppinfo}
\end{suppinfo}

\bibliography{references}

\includepdf[pages=-]{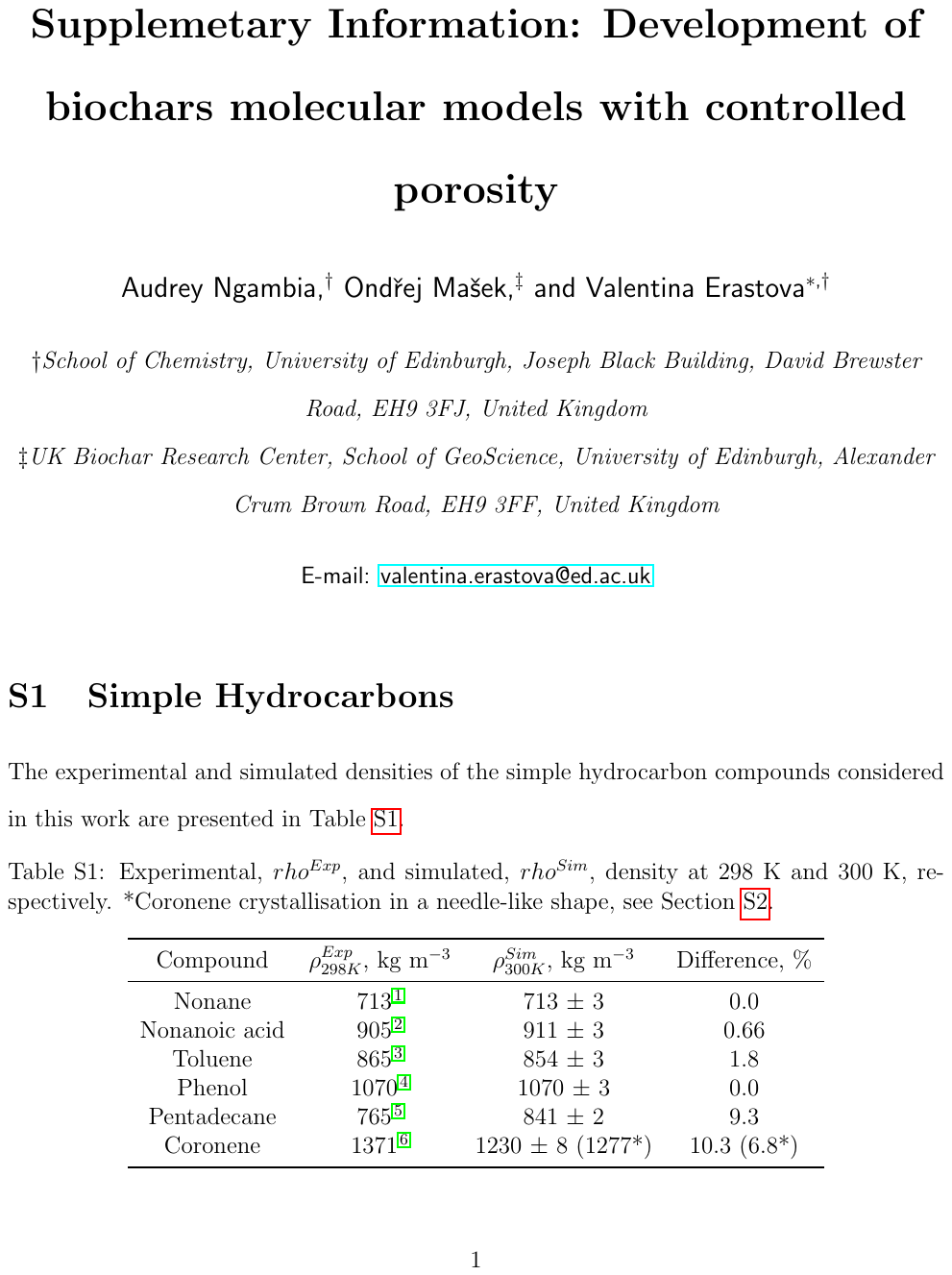}

\end{document}



\section{Simple Hydrocarbons}
The experimental and simulated densities of the simple hydrocarbon compounds considered in this work are presented in Table \ref{tab:Dens}.

\begin{table}[hbt!]
    \centering
    \begin{tabular}{cccc}
    \toprule
Compound & $\rho^{Exp}_{298 K}$, kg m$^{-3}$ &  $\rho^{Sim}_{300 K}$, kg m$^{-3}$ & Difference, \%\\
    \midrule
       Nonane  & 713 \cite{ramos2011experimental} &  713 $\pm$ 3  & 0.0   \\
       Nonanoic acid  & 905 \cite{wypych2015databook} & 911 $\pm$ 3 & 0.66  \\
       Toluene  & 865 \cite{tenorio2023experimental} & 854 $\pm$ 3 &   1.8 \\
       Phenol  & 1070 \cite{cole1960density} & 1070 $\pm$ 3 &  0.0 \\
       Pentadecane  & 765 \cite{prak2023n} & 841 $\pm$ 2 & 9.3   \\
       Coronene  & 1371\cite{haynes2016crc} & 1230 $\pm$ 8 (1277*) &   10.3 (6.8*) \\
    \bottomrule
    \end{tabular}
    \caption{Experimental, $rho^{Exp}$, and simulated, $rho^{Sim}$, density at 298 K and 300 K, respectively. *Coronene crystallisation in a needle-like shape, see Section \ref{Ch:Coronene}.}
    \label{tab:Dens}
\end{table}

\begin{figure}[hbt!]
        \centering
        \includegraphics[width=\linewidth]{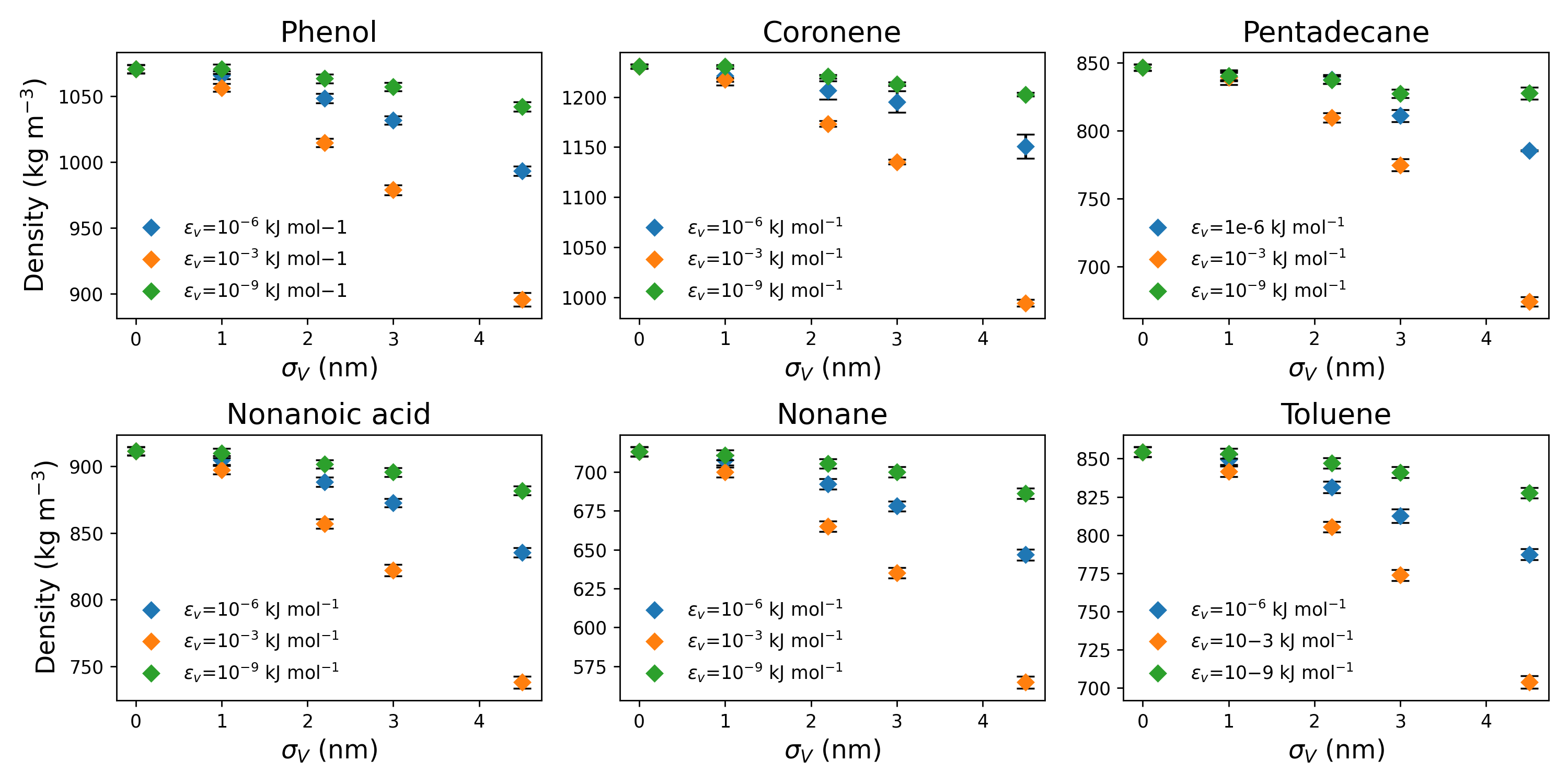}
        \caption{Change in density with virtual atom $\sigma$ and $\epsilon$.}
        \label{fig:10}
    \end{figure}

\break

\section{Coronene}
\label{Ch:Coronene}

\begin{figure}[hbt!]
    \centering
    \subfloat[Coronene needle-shape crystal.]
    {
        \label{fig:coro_needle}
        \includegraphics[width=0.7\linewidth]{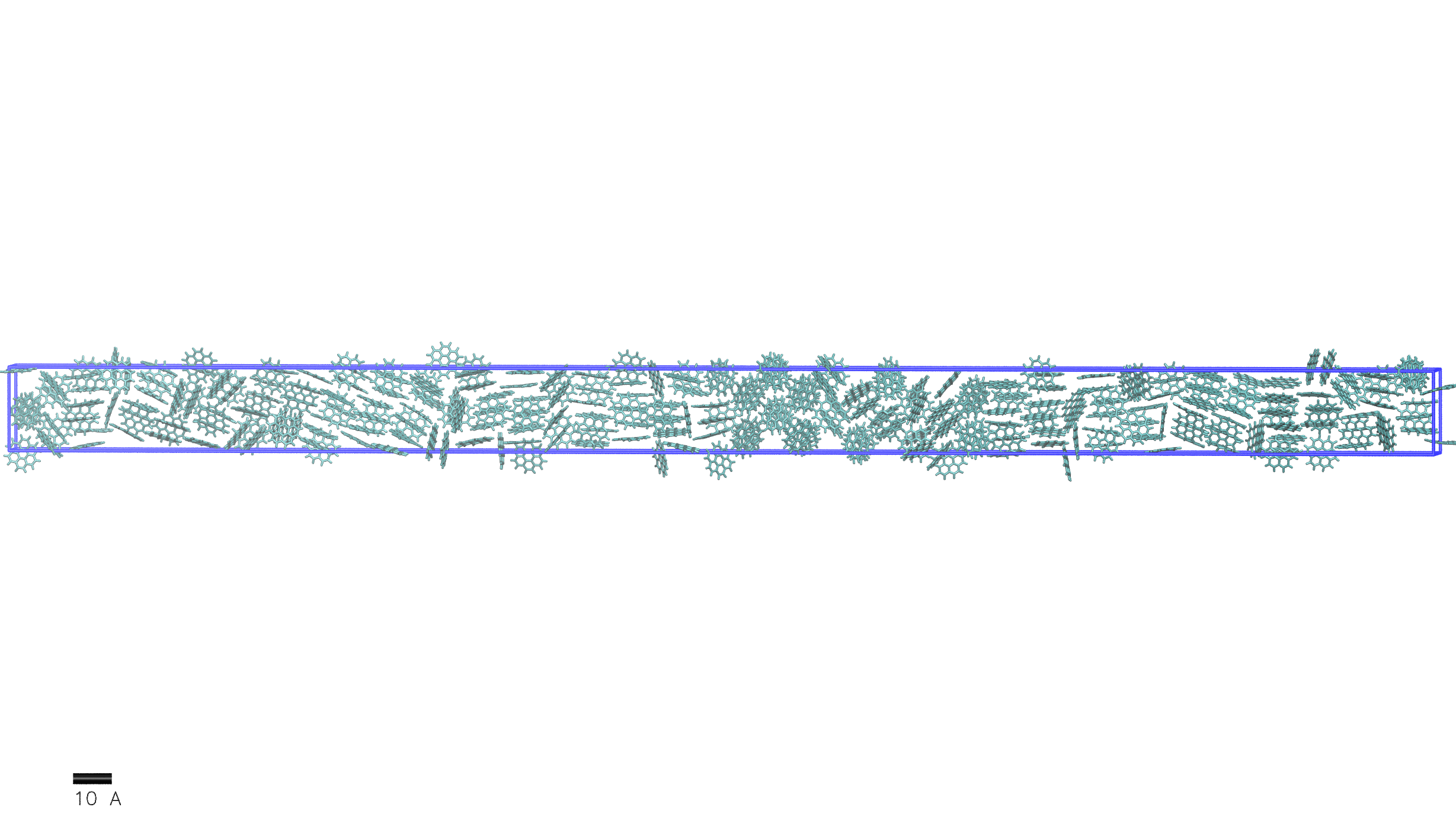}
    }\hfill
      \newline
        \subfloat[Coronene semi-amorphous crystallisation.] 
        {
        \label{fig:coronene}
        \includegraphics[width=0.5\textwidth]{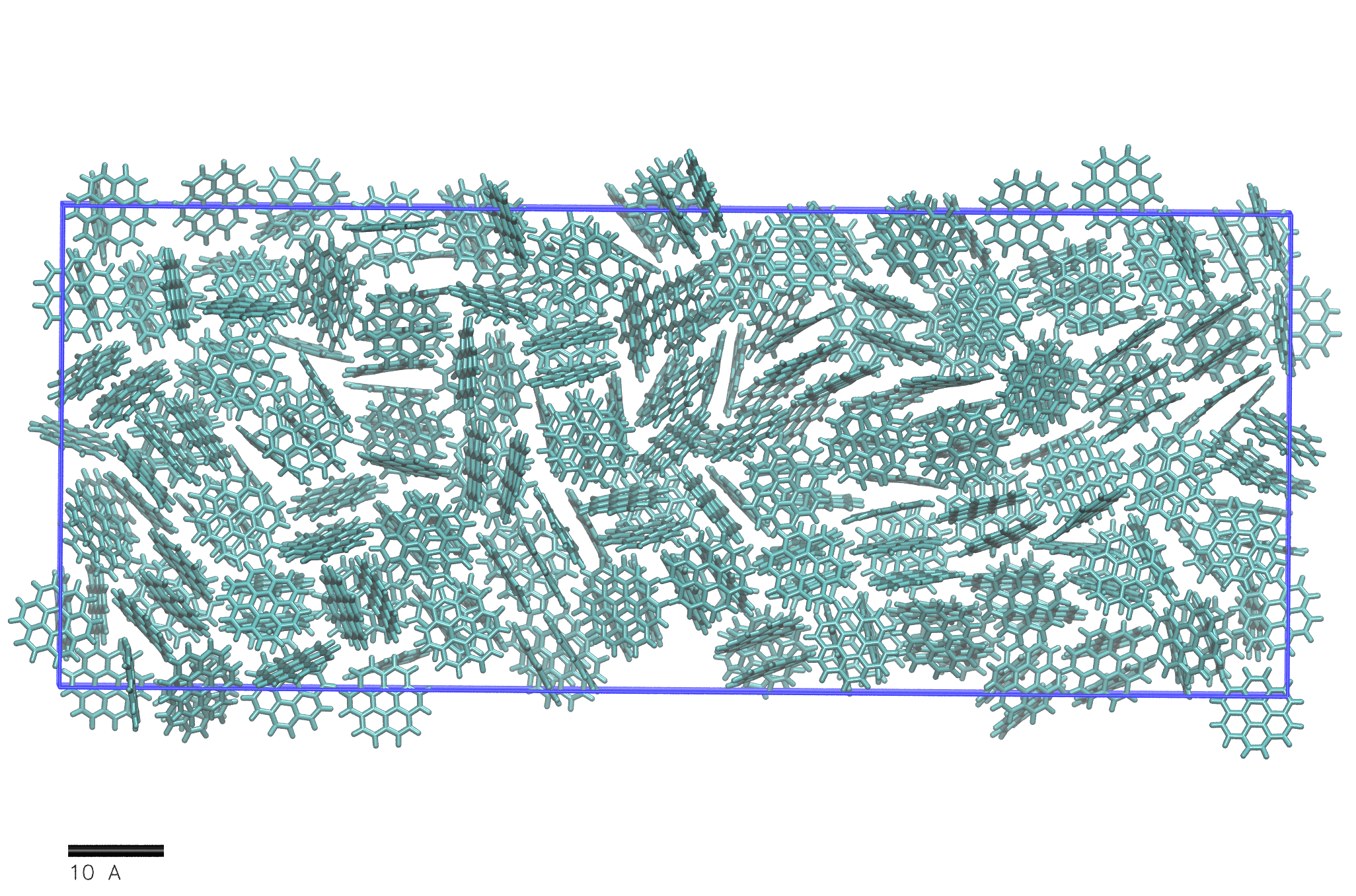}
    }\hfill
    
        \caption{Coronene crystallisation upon cooling from high (1000 K) to low (300 K) temperatures.}
    \label{fig:Coronene_crystal}
\end{figure}

Coronene adopts a needle-like crystal structure with a monoclinic-herringbone molecular arrangement. 
This means that within an MD simulation, a formation of crystal from liquid (cubic simulation box) will result in a rapid elongation of one of the axis and, consequently, a reduction of the others. 
Practically, this means that in order to obtain a true needle-like structure, one of axises of our simulation box will become smaller than the cut-off radius, requiring us to stop the simulation and double up the system in that direction. Combined with the already lengthy simulations, the increased atomic numbers of these needle-like systems create a significant resource demand. 
Figure \ref{fig:coro_needle} shows the needle-like shape of coronene, that was obtained during our simulation testing the force field. We note that the obtained crystal is still not ideal, with some structural defects, which is evident by an underestimated density by ~6.8\%. The crystallisation of solids in MD is a difficult task in itself, and not a goal of this work. Instead, since we were interested in creating a simulation protocol suitable for the test of VA interactions with various condensed organic moieties, we settled being able to create a semi-amourphous coronene, as shown on Figure \ref{fig:coronene}.  

\break

\section{Changes in volume with VA removal}

\begin{figure}[hbt!]
    \centering
    \subfloat[Pore volume change in BCMA models before and after removal of VAs.]
    {
        \label{fig:bcma_pore_stability}
        \includegraphics[width=0.8\linewidth]{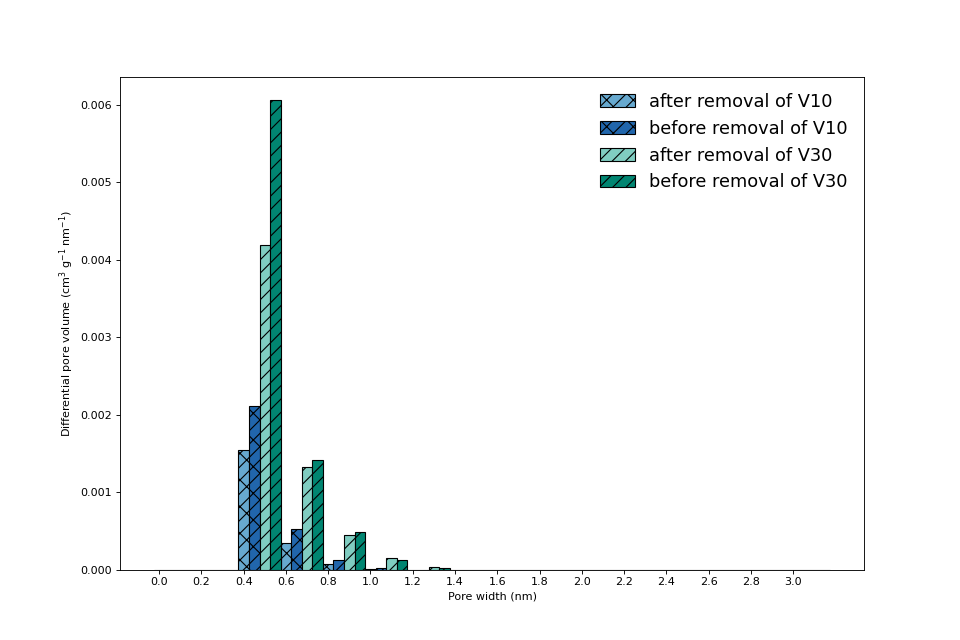}
    }\hfill
      \newline
        \subfloat[Pore volume change in BCMB models before and after removal of VAs.] 
        {
        \label{fig:bcmb_pore_stability}
        \includegraphics[width=0.8\linewidth]{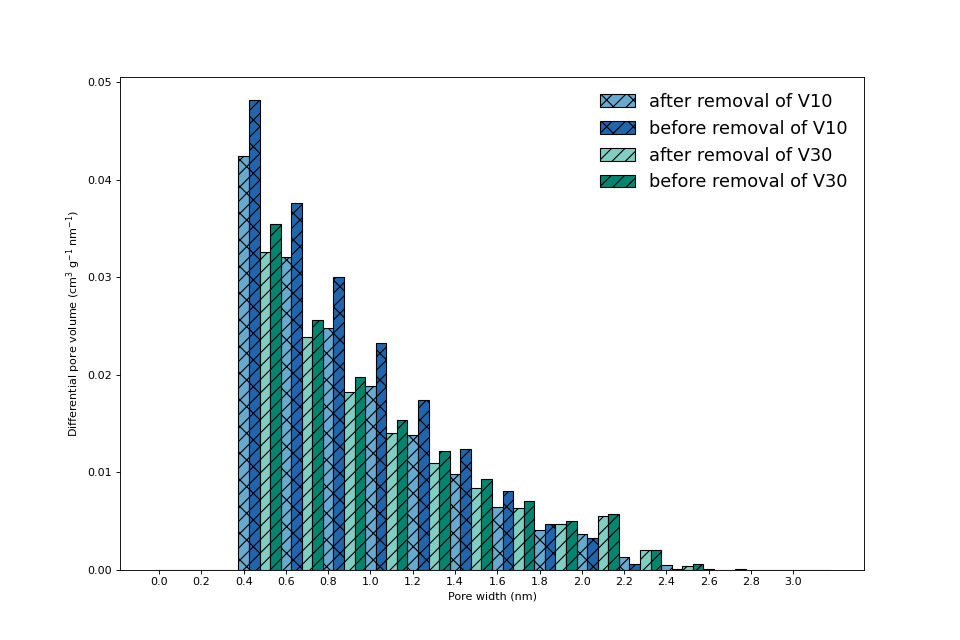}
    }\hfill
    
        \caption{Pore stability of BCMA and BCMB biochar model created with virtual atoms.}
    \label{fig:Pore_stability}
\end{figure}

\bibliography{references}